\documentstyle[12pt,epsfig]{article}
\def\bge{\begin{equation}}
\def\ene{\end{equation}}
\def\bg{\begin{eqnarray}}
\def\en{\end{eqnarray}}
\def\nn{\nonumber}

\begin{document}
\begin{titlepage}
\title{The quark-meson coupling model for \\
$\Lambda$, $\Sigma$ and $\Xi$ hypernuclei}  
\author{\\ 
K. Tsushima$^1$~\thanks{ktsushim@physics.adelaide.edu.au}~,
K. Saito$^2$~\thanks{ksaito@nucl.phys.tohoku.ac.jp}~, 
J. Haidenbauer$^3$~\thanks{j.haidenbauer@fz-juelich.de}~, and
A. W. Thomas$^1$~\thanks{athomas@physics.adelaide.edu.au} \\ \\
$^1$Department of Physics and Mathematical Physics \\
and \\
Special Research Center for the Subatomic Structure of Matter, \\ 
University of Adelaide, SA 5005, Australia \\ \\
$^2$Physics Division, Tohoku College of Pharmacy,  
Sendai 981, Japan \\ \\
$^3$Forschungszentrum J\"{u}lich, IKP, D-52425 J\"{u}lich, Germany \\ \\ 
}
\maketitle
\vspace{-14cm}
\hfill ADP-97-26/T261
\vspace{14cm}
\begin{abstract}
The quark-meson coupling (QMC) model, which has been successfully used to
describe the properties of both infinite nuclear matter and finite nuclei,
is applied to a systematic study of $\Lambda, \Sigma$ and $\Xi$ hypernuclei.
Assumptions made in the present study are, 
(i) the (self-consistent) exchanged scalar, and vector, mesons couple only 
to the u and d quarks, and (ii) an SU(6) valence 
quark model for the bound nucleons and 
hyperon. The model automatically 
leads to a very weak spin-orbit interaction for the $\Lambda$ in a
hypernucleus.
Effects of the Pauli blocking at the quark level, particularly  
in the open, coupled, $\Sigma N - \Lambda N$ channel (strong conversion),  
is also taken into account in a phenomenological way.
\end{abstract}
\end{titlepage}
%
%
\section{Introduction}

In earlier work~\cite{finite0,finite1,finite2} 
we addressed the question of whether quarks play 
an important role in finite nuclei.
This involved quantitative investigations of the properties of
closed-shell nuclei from $^{16}$O to $^{208}$Pb~\cite{finite0,finite1},
as well as the effective mass of the $\rho$ meson formed in 
light nuclei~\cite{finite2}. These calculations were performed 
within the quark-meson coupling (QMC) model, originally
suggested by Guichon~\cite{gui}, and with an extended version~\cite{finite2}, 
which treats also the $\omega$ and $\rho$ 
meson mass variations self-consistently in the medium. 

In the QMC model, the interactions between hadrons 
are mediated by the exchange of scalar ($\sigma$) and vector 
($\omega$ and $\rho$) mesons self-consistently coupled to the quarks within 
those hadrons. 
Within the model it has proven possible to successfully describe 
the properties of infinite nuclear matter~\cite{gui,matter}    
and also finite nuclei~\cite{finite0,finite1,finite2}.
Blunden and Miller~\cite{blu}, and Jin and Jennings~\cite{jin} 
have made similar studies based on the QMC model, 
and some phenomenological extensions.
One of the most attractive features of the QMC model is 
that it does not involve much in the way of additional complications to
Quantum Hadrodynamics (QHD)~\cite{qhd}. 
Furthermore, it produces a reasonable value for 
the nuclear incompressibility~\cite{finite0}-\cite{jin}.

Here we apply the QMC model~\cite{finite0,finite1}  
to a systematic study of the properties 
of $\Lambda$, $\Sigma$ and $\Xi$ hypernuclei. 
Some of the initial results for $\Lambda$ hypernuclei were 
reported already~\cite{hyper}.
One of the purposes of the present article is to present  
the spin-orbit potentials for the hyperon 
calculated self-consistently with an explicit quark structure 
for the bound hyperon in the QMC model.
These spin-orbit potentials generally contain anomalous contributions
from the quarks due to the finite size of the hyperon. 

Within the Born-Oppenheimer approximation, one can derive equations of 
motion for a hypernucleus in the QMC model~\cite{hyper} 
in the same way as has been 
done for normal nuclei~\cite{finite0,finite1}. Such an approach may provide 
us with important information about 
the hyperon-nucleon interaction, the deep nuclear interior and 
a possible manifestation of the quark degrees of freedom 
via the Pauli principle at the quark level~\cite{bou}-\cite{yam}.
Below, we will briefly review the present situation 
regarding $\Lambda$ hypernuclei first,  
because they have been well studied, both experimentally 
and theoretically.

As an example, the very weak spin-orbit interaction for 
$\Lambda$ hypernuclei, which had been 
phenomenologically suggested by Bouyssy and H\"{u}fner~\cite{bou}, 
was first explained by Brockman 
and Weise~\cite{bro} in a relativistic Hartree model,
and directly confirmed later by experiment~\cite{bru}.
However, a very strong SU(3) breaking effect was required to 
achieve this small spin-orbit force. 
An explanation in terms of quark and gluon dynamics 
was made by Pirner~\cite{pir}.
Alternatively, Noble~\cite{nob} showed that this small 
spin-orbit force could be also realized without 
any large breaking of SU(3) symmetry,  
if an $\omega \Lambda \Lambda$ tensor coupling was introduced, 
analogous to the anomalous magnetic moment of the $\Lambda$. 
However, Dover and Gal~\cite{dov} questioned whether the tensor 
coupling of the $\omega$ meson to the $\Lambda$ could 
be related to the anomalous magnetic moment,
because the spin of $\Lambda$ is entirely carried 
by the s quark in a naive SU(6) valence quark model, and this s quark  
couples exclusively to the $\phi$ meson according to the OZI rule.
Later, Jennings~\cite{jen} pointed out that within Dirac phenomenology 
the tensor coupling of the $\omega$ meson to the $\Lambda$ 
could be introduced in such a way as to guarantee that the direct $\omega$
coupling to the spin of the $\Lambda$ was zero -- as one would expect
from a simple quark model. The resulting spin-orbit force agrees with
the result obtained in the present model~\cite{hyper}, 
on the basis of an explicit
treatment of the quark structure of the $\Lambda$ moving in vector and
scalar fields that vary in space.

As a second example, it has also been discovered that there is an overbinding 
problem in the light $\Lambda$ hypernuclei, and 
the existence of a repulsive core or the necessity of a 
repulsive three-body force 
have been suggested to overcome the problem~\cite{bod}.
The origin of this overbinding, 
which could not be explained easily in terms of traditional nuclear 
physics, was ascribed to the Pauli principle at the quark level 
by Hungerford and Biedenharn~\cite{hun}.
Investigations of this repulsive core in the $\Lambda$-nucleus system  
have been made by Takeuchi and Shimizu, and others~\cite{tak,str}, 
based on a nonrelativistic quark model.

In addition to the investigations of $\Lambda$ hypernuclei, 
many studies of $\Sigma$ and $\Xi$ hypernuclei have been 
also made~\cite{dov},\cite{bro2}-\cite{dovy},\cite{yyam}-\cite{yam}.
In particular, experiments on $\Sigma$ hypernuclei  
performed at CERN~\cite{ber}, 
Brookhaven~\cite{pie} and KEK~\cite{yam}, confirmed the existence 
of $\Sigma$ hypernuclei. The former two experiments were done 
using $K^-$ beams in so called $\lq\lq$in-flight kinematics", 
while the latter was done using stopped kaons.
These experiments revealed an interesting feature of 
$\Sigma$ hypernuclei, namely, the widths of the $\Sigma$ in a hypernucleus 
seems to be narrower than the naive expectations due to the strong 
$\Sigma N - \Lambda N$ conversion (channel coupling).
As for the $\Xi$ hypernuclei, although not so many 
experiments on the double strangeness hypernuclei (S$\,\,= -2$) 
have been performed so far~\cite{dan}-\cite{ima}, there is a 
renewed interest in connection with the H dibaryon suggested 
by Jaffe~\cite{jaf}, based on quark degrees of 
freedom -- using the MIT bag model.
Thus, we expect that more experiments   
on the $\Sigma$ and $\Xi$ hypernuclei,  
with higher statistics and precision, will be performed in the near future.  

In the light of these investigations, it now seems appropriate to
investigate the (heavier) hypernuclear systems, 
$\Lambda$, $\Sigma$ and $\Xi$ hypernuclei quantitatively, 
using a microscopic model based on quark degrees of freedom.
For this purpose, the QMC model (which is built explicitly on 
quark degrees of freedom) 
seems ideally suited, because it has already been shown to describe 
the properties of finite nuclei quantitatively.
The present investigation is one of the natural extensions of the study 
of quark degrees of freedom in finite nuclei.

The organization of the paper is as follows. In Section 2, the relativistic 
formulation of the hypernuclear system in QMC will be 
explained. A mean-field Lagrangian density and equations of motion will 
be derived in Section 2.1. Hyperons in infinite 
nuclear matter will be discussed in 
Section 2.2, while (finite) hypernuclei will be treated in Section 2.3.
In Section 3, the spin-orbit potential for the hyperon 
in the QMC model will be discussed. 
Some results for hypernuclei in the QMC model alone will be given 
in Section 4. In Section 5, modifications necessary for a more realistic 
calculation in addition to the present version 
of the QMC model will be discussed.
Effects of the Pauli blocking 
at the quark level in the $\Lambda N - \Sigma N$ 
(and $\Xi N - \Lambda \Lambda$) coupled channel, 
will be considered in specific ways 
at the hadronic level in Sections 5.1, and 5.2, respectively.
The results with including these effects discussed in 
Sections 5.1 and 5.2 will be given in Section 5.3.
Finally, Section 6 will be devoted for summary and discussion.

%
%
\section{Hypernuclei in the QMC model}

In this Section, we will derive mean-field equations of motion for a
hypernucleus, as well as consider a hyperon in nuclear matter.

\subsection{Mean-field equations of motion}
Using the Born-Oppenheimer approximation one can derive mean-field equations 
of motion for a hypernucleus in which the quasi-particles moving 
in single-particle orbits are three-quark clusters with the quantum numbers 
of a hyperon or a nucleon. One can then construct a relativistic 
Lagrangian density at the hadronic level~\cite{finite0,finite1,hyper}, 
similar to that obtained in QHD~\cite{qhd}, 
which produces the same equations of motion 
when expanded to the same order in velocity, $v$: 
\begin{eqnarray}
{\cal L}^{HY}_{QMC} &=& {\cal L}_{QMC} + {\cal L}^Y_{QMC}, 
\nn \\
{\cal L}_{QMC} &=&  \overline{\psi}_N(\vec{r}) 
\left[ i \gamma \cdot \partial
- M_N^{\star}(\sigma) - (\, g_\omega \omega(\vec{r}) 
+ g_\rho \frac{\tau^N_3}{2} b(\vec{r}) 
+ \frac{e}{2} (1+\tau^N_3) A(\vec{r}) \,) \gamma_0 
\right] \psi_N(\vec{r}) \quad \nn \\
  &-& \frac{1}{2}[ (\nabla \sigma(\vec{r}))^2 +
m_{\sigma}^2 \sigma(\vec{r})^2 ]
+ \frac{1}{2}[ (\nabla \omega(\vec{r}))^2 + m_{\omega}^2
\omega(\vec{r})^2 ] \nn \\
 &+& \frac{1}{2}[ (\nabla b(\vec{r}))^2 + m_{\rho}^2 b(\vec{r})^2 ]
+ \frac{1}{2} (\nabla A(\vec{r}))^2, \nn \\
{\cal L}^Y_{QMC} &=& \sum_{Y=\Lambda,\Sigma,\Xi} \overline{\psi}_Y(\vec{r}) 
\left[ i \gamma \cdot \partial
- M_Y^{\star}(\sigma)
- (\, g^Y_\omega \omega(\vec{r}) 
+ g^Y_\rho I^Y_3 b(\vec{r}) 
+ e Q_Y A(\vec{r}) \,) \gamma_0 
\right] \psi_Y(\vec{r}), \qquad \label{lagrangian}
\end{eqnarray}
where $\psi_N(\vec{r})$ ($\psi_Y(\vec{r})$)  
and $b(\vec{r})$ are respectively the
nucleon (hyperon) and the $\rho$ 
meson (the time component in the third direction of
isospin) fields, while $m_\sigma$, $m_\omega$ and $m_{\rho}$ are 
the masses of the $\sigma$, $\omega$ and $\rho$ mesons.
$g_\omega$ and $g_{\rho}$ are the $\omega$-N and $\rho$-N
coupling constants which are related to the corresponding 
(u,d)-quark-$\omega$, $g_\omega^q$, and 
(u,d)-quark-$\rho$, $g_\rho^q$, coupling constants as
$g_\omega = 3 g_\omega^q$ and 
$g_\rho = g_\rho^q$~\cite{finite0,finite1}.


In an approximation where the $\sigma$, $\omega$ and $\rho$ mesons couple
only to the u and d quarks (ideal mixing of the $\omega$ and $\phi$ 
mesons, and the OZI rule are assumed),
the coupling constants in the hyperon sector
are obtained as $g^Y_\omega = (n_0/3) g_\omega$, and 
$g^Y_\rho = g_\rho = g_\rho^q$, with $n_0$ being the total number of
valence u and d quarks in the hyperon Y. $I^Y_3$ and $Q_Y$
are the third component of the hyperon isospin operator and its electric 
charge in units of the proton charge, $e$, respectively. 
The field dependent $\sigma$-N and $\sigma$-Y
coupling strengths predicted by the QMC model,  
$g_\sigma(\sigma)$ and  $g^Y_\sigma(\sigma)$, 
related to the Lagrangian density 
Eq.~(\ref{lagrangian}) at the hadronic level, are defined by: 
\bg
M_N^{\star}(\sigma) &\equiv& M_N - g_\sigma(\sigma)
\sigma(\vec{r}) ,  \\
M_Y^{\star}(\sigma) &\equiv& M_Y - g^Y_\sigma(\sigma)
\sigma(\vec{r}) , \label{coupny}
\en
where $M_N$ ($M_Y$) is the free nucleon (hyperon) mass.
Note that the dependence of these coupling strengths on the applied
scalar field must be calculated self-consistently within the quark
model. Hence, unlike QHD, even though 
$g^Y_\sigma(\sigma) / g_\sigma(\sigma)$ may be
2/3 in free space ($\sigma = 0$)\footnote{Strictly, this is true  
only when the bag radii of nucleon and hyperon are exactly the same
in the present model. 
See Eq.~(\ref{cconst}), below.} 
, this will not necessarily  be the case in
nuclear matter.
More explicit expressions for $g^Y_\sigma(\sigma)$ 
and $g_\sigma(\sigma)$ will be given later. From 
the Lagrangian density 
Eq.~(\ref{lagrangian}), one gets a set of 
equations of motion for the hypernuclear system: 
\begin{eqnarray}
& &[i\gamma \cdot \partial -M^{\star}_N(\sigma)-
(\, g_\omega \omega(\vec{r}) + g_\rho \frac{\tau^N_3}{2} b(\vec{r}) 
 + \frac{e}{2} (1+\tau^N_3) A(\vec{r}) \,) 
\gamma_0 ] \psi_N(\vec{r}) = 0, \label{eqdiracn}\\
& &[i\gamma \cdot \partial - M^{\star}_Y(\sigma)-
(\, g^Y_\omega \omega(\vec{r}) + g_\rho I^Y_3 b(\vec{r}) 
+ e Q_Y A(\vec{r}) \,) 
\gamma_0 ] \psi_Y(\vec{r}) = 0, \label{eqdiracy}\\
& &(-\nabla^2_r+m^2_\sigma)\sigma(\vec{r}) = 
- [\frac{\partial M_N^{\star}(\sigma)}{\partial \sigma}]\rho_s(\vec{r})  
- [\frac{\partial M_Y^{\star}(\sigma)}{\partial \sigma}]\rho^Y_s(\vec{r}),
\nn \\
& & \hspace{7.5em} \equiv g_\sigma C_N(\sigma) \rho_s(\vec{r})
    + g^Y_\sigma C_Y(\sigma) \rho^Y_s(\vec{r}) , \label{eqsigma}\\
& &(-\nabla^2_r+m^2_\omega) \omega(\vec{r}) =
g_\omega \rho_B(\vec{r}) + g^Y_\omega 
\rho^Y_B(\vec{r}) ,\label{eqomega}\\
& &(-\nabla^2_r+m^2_\rho) b(\vec{r}) =
\frac{g_\rho}{2}\rho_3(\vec{r}) + g^Y_\rho I^Y_3 \rho^Y_B(\vec{r}),  
 \label{eqrho}\\
& &(-\nabla^2_r) A(\vec{r}) = 
e \rho_p(\vec{r}) 
+ e Q_Y \rho^Y_B(\vec{r}) ,\label{eqcoulomb}
\end{eqnarray}
where, $\rho_s(\vec{r})$ ($\rho^Y_s(\vec{r})$), $\rho_B(\vec{r})$ 
($\rho^Y_B(\vec{r})$), $\rho_3(\vec{r})$ and 
$\rho_p(\vec{r})$ are the scalar, baryon, third component of isovector,   
and proton densities at the position $\vec{r}$ in 
the hypernucleus~\cite{finite1,finite2}.    
On the right hand side of Eq.~(\ref{eqsigma}),
a new, and characteristic feature of QMC beyond QHD~\cite{qhd,ruf,coo}
appears, namely,
$- \frac{\partial M_N^{\star}(\sigma)}{\partial \sigma} = 
g_\sigma C_N(\sigma)$ and 
$- \frac{\partial M_Y^{\star}(\sigma)}{\partial \sigma} = 
g^Y_\sigma C_Y(\sigma)$, where $g_\sigma \equiv g_\sigma (\sigma=0)$ and 
$g^Y_\sigma \equiv g^Y_\sigma (\sigma=0)$. 
The effective mass for the hyperon Y is defined by

\begin{equation}
M_Y^{\star}(\sigma) =
\frac{n_0\Omega^{\star}(\sigma) 
+ (3-n_0)\Omega^{\star}_s - z_Y}{R_Y}
+ \frac{4}{3}\pi (R_Y)^3 B ,\label{bagmass}\\
\end{equation}
with the MIT bag model quantities~\cite{finite0,finite1} 
\begin{eqnarray}
S_Y(\sigma) &=& \frac{\Omega^{\star}(\sigma)/2 
+ m_{u,d}^{\star}(\sigma)R_Y(\Omega^{\star}(\sigma)-1)}
{\Omega^{\star}(\sigma)(\Omega^{\star}(\sigma)-1) 
+ m_{u,d}^{\star}(\sigma)R_Y/2}, \label{ssigma}\\ 
C_Y(\sigma) &=& S_Y(\sigma)/S_Y(0), \label{csigma}\\
g^Y_{\sigma} &\equiv& n_0 g_{\sigma}^q S_Y(0) 
= \frac{n_0}{3} g_\sigma S_Y(0)/S_N(0) 
\equiv \frac{n_0}{3}g_\sigma \Gamma_{Y/N}, \label{cconst}\\
\Omega^{\star}(\sigma) &=& \sqrt{x^2 + (R_Ym_{u,d}^{\star}(\sigma))^2}, 
\qquad \Omega_s^{\star} = \sqrt{x_s^2 + (R_Ym_s)^2},
\label{omega}\\
m_{u,d}^{\star}(\sigma) &=& m_{u,d} - g_{\sigma}^q \sigma (\vec{r}), 
\label{qmass}
\end{eqnarray}
such $R_Y$ that satisfies 
\begin{equation}
\left. \frac{dM_Y^\star (\sigma)}{d R_Y}
\right|_{R_Y=R_Y^\star} = 0. 
\label{bagradius}
\end{equation}
The density dependent $\sigma$-$Y$ coupling constant is defined by 
\begin{equation}
\frac{\partial M_Y^{\star}(\sigma)}{\partial \sigma}
= - n_0 g_{\sigma}^q \int_{bag} d\vec{y} 
\ {\overline \psi}_{u,d}(\vec{y}) \psi_{u,d}(\vec{y})
\equiv - n_0 g_{\sigma}^q S_Y(\sigma) = - \frac{\partial}{\partial \sigma}
\left[ g^Y_\sigma(\sigma) \sigma \right].
\label{dmds}
\end{equation}
For the nucleon, replace the suffix $Y$ by $N$ 
in Eqs.~(\ref{bagmass}) - (\ref{dmds}) and set $n_0 = 3$.
Here, $z_Y$, $B$, $x$, $x_s$ and $m_{u,d,s}$ are the parameters 
for the sum of the c.m.
and gluon fluctuation effects,
bag pressure, lowest eigenvalues for the (u, d) and s quarks, respectively, 
and the corresponding current quark masses with $m_u = m_d$.
$z_N$ and $B$ ($z_Y$) are fixed by fitting the nucleon (hyperon) mass 
in free space. 
The bag parameters calculated and chosen for the present study are,
$R_N = 0.8$ fm (in free space),
$B = (170$ MeV$)^4$, $m_u = m_d = 5$ MeV and $m_s = 250$ MeV. 
Note that due to the equilibrium condition Eq.~(\ref{bagradius}), 
which is applied also to the free nucleonic vacuum, 
the bag radius for each hyperon in free space is slightly different. 
The parameters $z_{N,Y}$ and the bag radii, $R_{N,Y}$, obtained  
by fitting to their physical masses in free space are listed in 
Table~\ref{bagparam}.

%
\begin{table}[htbp]
\begin{center}
\caption{The bag parameters, $z_{N,Y}$, 
the bag radii in free space, $R_{N,Y}$ and 
the physical baryon masses fitted in free space.
They are obtained with the bag pressure, $B = (170$ MeV$)^4$, current 
quark masses, $m_u = m_d = 5$ MeV and $m_s = 250$ MeV.}
\label{bagparam}
\begin{tabular}[t]{c|ccc}
\hline
&mass (MeV)  &$z_{N,Y}$ &$R_{N,Y}$ (fm)\\ 
\hline 
$N$       &939.0  &3.295 &0.800\\
$\Lambda$ &1115.7 &3.131 &0.806\\
$\Sigma$  &1193.1 &2.810 &0.827\\
$\Xi$     &1318.1 &2.860 &0.820\\
\end{tabular}
\end{center}
\end{table}
%

The parameters associated with the u and d quarks are those found in 
our previous investigations~\cite{finite1}. 
The value for the mass of the s quark was 
chosen to be $250$ MeV. 
We note that the final results are insensitive 
to this parameter. 
The value for $\Gamma_{Y/N}$ in Eq.~(\ref{cconst}) 
turned out to be almost unity for all hyperons 
even though $R_N \ne R_Y$ (see also Eqs.~(\ref{ssigma})), 
so we can use $\Gamma_{Y/N} = 1$ in practice~\cite{finite2}. 
Although the bag radii in free space for the nucleon and 
hyperons are slightly different ($R_N \neq R_Y$), 
they will be generically denoted by 
$R_B = 0.8$ fm in Figs.~\ref{csigmaf} and~\ref{delmassf}.

At the hadronic level, the entire information 
on the quark dynamics is condensed into the effective coupling 
$C_{N,Y}(\sigma)$ of Eq.~(\ref{eqsigma}), or Eq.~(\ref{csigma}).
Furthermore, when $C_{N,Y}(\sigma) = 1$, which corresponds to 
a structureless nucleon or hyperon, the equations of motion
given by Eqs.~(\ref{eqdiracn})-(\ref{eqcoulomb})
can be identified with those derived from QHD~\cite{mar,ruf,coo}, 
except for the terms arising from the tensor coupling and the non-linear 
scalar field interaction introduced beyond naive QHD.

%
\subsection{Nuclear matter limit}

\begin{figure}[hbt]
\begin{center}
\epsfig{file=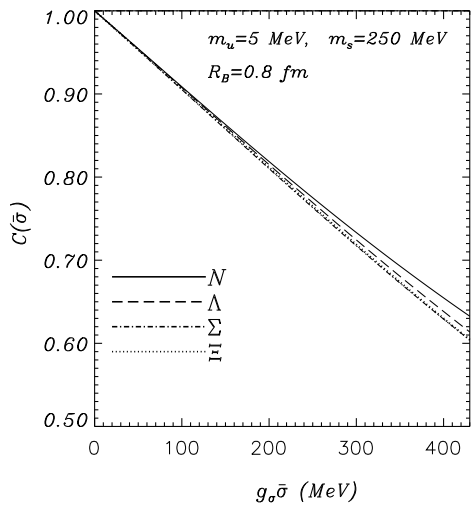,height=7cm}
\caption{$C(\overline{\sigma})$ 
for nucleon and hyperons in nuclear matter.}
\label{csigmaf}
\end{center}
\end{figure}

Here, we consider a hyperon in nuclear matter. In this limit the meson fields 
become constants, and we denote the mean-value of the $\sigma$ field as
$\overline{\sigma}$.
The self-consistency condition for the $\sigma$ field, $\overline{\sigma}$,  
and effective nucleon mass, $M_N^\star$, 
is given by~\cite{finite0,finite1,matter} 
\bg
\overline{\sigma}&=&\frac{g_\sigma }{m_\sigma^2}C_N(\overline{\sigma})
\frac{4}{(2\pi)^3}\int d\vec{k} \theta (k_F - k)
\frac{M_N^{\star}(\overline{\sigma})}
{\sqrt{M_N^{\star 2}(\overline{\sigma})+\vec{k}^2}} , \label{scc}
\en
where $g_{\sigma} = (3 g^q_\sigma S_N(0))$, $k_F$ is the 
Fermi momentum, 
and $C_N(\overline{\sigma})$ is now the constant value of $C_N$ in the
scalar field.
Note that $M_N^\star (\overline{\sigma})$,  
in Eq.~(\ref{scc}), must be calculated 
self-consistently by the MIT bag model, through 
Eqs.~(\ref{bagmass}) - (\ref{dmds}).
This self-consistency equation for $\overline{\sigma}$
is the same as that in QHD, except that in the latter model one has
$C_N(\overline{\sigma})=1$~\cite{qhd}. 
By using the obtained mean field value, $\overline{\sigma}$, 
the corresponding quantity for the hyperon Y, 
$C_Y(\overline{\sigma})$, can be also calculated using 
Eqs.~(\ref{ssigma}) and~(\ref{csigma}), where the effect of a single 
hyperon on the 
mean field value, $\overline{\sigma}$, in infinite nuclear 
matter can be neglected.
The calculated values of $C_{N,Y}(\overline{\sigma})$ for 
the nucleon and hyperons 
are displayed in Fig.~\ref{csigmaf}.
%
%
%
%
It has been found that the 
function $C_{N,Y}(\overline{\sigma})$ can be parametrized as a linear 
form in the $\sigma$ field, $g_{\sigma}\overline{\sigma}$, for practical 
calculations~\cite{finite0,finite1,finite2} by 
\begin{equation}
C_{N,Y} (\overline{\sigma}) = 1 - a_{N,Y} 
\times (g_{\sigma} \overline{\sigma}),
\label{cynsigma}
\end{equation}
which can be also seen from Fig.~\ref{csigmaf},
where $g_{\sigma} \overline{\sigma} 
= (3 g^q_\sigma S_N(0)) \,\overline{\sigma}$ in MeV. 
The values of $a_{N,Y}$ for nucleon and hyperons are given 
in Table~\ref{slope}.
%
\begin{table}[htbp]
\begin{center}
\caption{The slope parameters, $a_N$ and $a_Y$.}
\label{slope}
\begin{tabular}[t]{c|c}
\hline
&$\times 10^{-4}$ MeV$^{-1}$\\ 
\hline 
$a_N$       &8.8\\
$a_\Lambda$ &9.3\\
$a_\Sigma$  &9.5\\
$a_\Xi$     &9.4\\
\end{tabular}
\end{center}
\end{table}
%
This parametrization works very well up to 
about three times normal nuclear density, $\rho_B \simeq 3 \rho_0$, 
with $\rho_0 \simeq 0.15$ fm$^{-3}$.
For the field strength, $g_\sigma \overline{\sigma}$, 
versus baryon density, see Ref.~\cite{finite0}.

\begin{figure}[hbt]
\begin{center}
\epsfig{file=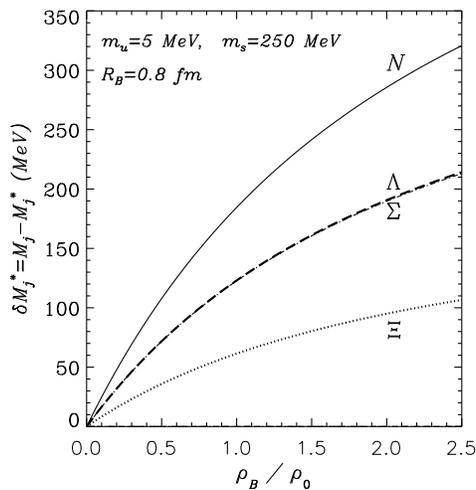,height=7cm}
\caption{Variation of the effective masses for nucleon and hyperons 
in nuclear matter, where normal nuclear matter density, 
$\rho_0$, is 0.15 fm$^{-3}$.}
\label{delmassf}
\end{center}
\end{figure}

Next, we define the variations of effective masses 
for the baryon $j\, (j = N, \Lambda, \Sigma, \Xi)$: 
\bge
\delta M_j^\star \equiv M_j - M_j^\star,
\label{delmasse}
\ene
which should be calculated self-consistently through the corresponding 
Eqs.~(\ref{bagmass})-(\ref{scc}). 
Calculated values for the variation of effective masses 
against baryon density are shown in Fig.~\ref{delmassf}.
%
%
%
For the effective mass, $M^\star_j$, one can write down explicit 
expressions using the parametrization for $C_{N,Y}(\overline{\sigma})$ 
of Eq.~(\ref{cynsigma}):
\bg
M^{\star}_N &\equiv& M_N - g_\sigma(\overline{\sigma}) \overline{\sigma}  
\simeq M_N - 
g_\sigma \left[ 1 - \frac{a_N}{2} (g_\sigma 
\overline{\sigma}) \right] \overline{\sigma}, \nn \\
M^{\star}_Y &\equiv& M_Y - g^Y_\sigma(\overline{\sigma}) \overline{\sigma}
\simeq M_Y - \frac{n_0}{3} 
g_\sigma \left[ 1 - \frac{a_Y}{2} (g_\sigma 
\overline{\sigma}) \right] \overline{\sigma}.
\label{mass} 
\en
Because the parameters $a_{N,Y}$ 
are almost the same in magnitudes (See Table~\ref{slope}), 
we find a simple scaling relation for the variation of 
effective masses~\cite{finite2}:
\bge
\frac{\delta M_\Lambda^\star}{\delta M_N^\star} \simeq 
\frac{\delta M_\Sigma^\star}{\delta M_N^\star} \simeq \frac{2}{3},\qquad
\frac{\delta M_\Xi^\star}{\delta M_N^\star} \simeq \frac{1}{3}.
\ene
This scaling relation is well demonstrated in Fig.~\ref{delmassf}.
It is worth noting that the ratio of the effective mass to 
the bare mass,  $M_j^\star /M_j$, does not decrease linearly 
as the baryon density increases in the QMC model, which is different from the 
relation adopted by Kuwabara and Hatsuda~\cite{hat}, based on a 
model of the QHD type.

%
\subsection{Hypernuclei}

The variation of the Lagrangian density Eq.~(\ref{lagrangian}) results in the
following equations for static, spherically symmetric nuclei plus 
one hyperon configuration (Hereafter, we will call this configuration 
simply a hypernucleus):
\bg
\frac{d^2}{dr^2} \sigma(r) + \frac{2}{r} \frac{d}{dr} \sigma(r)
    - m_\sigma^2 \sigma(r) &=& - g_\sigma C_N(\sigma(r)) \rho_s(r) 
                               - g^Y_\sigma C_Y(\sigma(r)) \rho^Y_s(r), \\ 
\frac{d^2}{dr^2} \omega(r) + \frac{2}{r} \frac{d}{dr} \omega(r)
    - m_\omega^2 \omega(r) &=& - g_\omega \rho_B(r) 
                               - g^Y_\omega \rho^Y_B(r), \\
\frac{d^2}{dr^2} b(r) + \frac{2}{r} \frac{d}{dr} b(r)
    - m_\rho^2 b(r) &=& - \frac{g_\rho}{2} \rho_3(r) 
                        - g^Y_\rho t_\beta \rho^Y_B(r), \\
\frac{d^2}{dr^2} A(r) + \frac{2}{r} \frac{d}{dr} A(r)
    &=& - e \rho_p(r) - eQ_Y \rho^Y_B(r), \\
\en
where
\bg
\rho_s(r)
&=& \sum_\alpha^{occ} d_\alpha(r)
(|G_\alpha(r)|^2 - |F_\alpha(r)|^2), \\
\rho^Y_s(r)
&=& d_\beta(r)
(|G_\beta(r)|^2 - |F_\beta(r)|^2), \\
\rho_B(r)
&=& \sum_\alpha^{occ} d_\alpha(r)
(|G_\alpha(r)|^2 + |F_\alpha(r)|^2), \\
\rho^Y_B(r)
&=& d_\beta(r) (|G_\beta(r)|^2 + |F_\beta(r)|^2), \\
\rho_3(r)
&=& \sum_\alpha^{occ}
d_\alpha(r) (-)^{t_\alpha -1/2}
(|G_\alpha(r)|^2 + |F_\alpha(r)|^2),  \\
\rho_p(r)
&=& \sum_\alpha^{occ} d_\alpha(r)
(t_\alpha + \frac{1}{2})
(|G_\alpha(r)|^2 + |F_\alpha(r)|^2),  
\en
with $d_\alpha(r)= (2j_\alpha+1)/4\pi r^2$, 
$d_\beta(r)= 1/4\pi r^2$ and
\bg
\frac{d}{dr} G_\alpha(r) + \frac{\kappa}{r} G_\alpha(r) -
\left[ \epsilon_\alpha - g_\omega \omega(r) - t_\alpha g_\rho b(r)
\right.
&-& \left. (t_\alpha + \frac{1}{2}) e A(r) + M_N \right. \nn \\
&-& \left. g_\sigma(\sigma(r)) \sigma(r) \right] F_\alpha(r) = 0,
\label{qwave1} \\
\frac{d}{dr} F_\alpha(r) - \frac{\kappa}{r} F_\alpha(r) +
\left[ \epsilon_\alpha - g_\omega \omega(r) - t_\alpha g_\rho b(r)
\right.
&-& \left. (t_\alpha + \frac{1}{2}) e A(r) - M_N \right. \nn \\
&+& \left. g_\sigma (\sigma(r)) \sigma (r) \right] G_\alpha(r) = 0,
\label{qwave2}
\en
\bg
\frac{d}{dr} G_\beta(r) + \frac{\kappa}{r} G_\beta(r) -
\left[ \epsilon_\beta - g^Y_\omega \omega(r) - t_\beta g^Y_\rho b(r)
\right.
&-& \left.  e Q_Y A(r) + M_Y \right. \nn \\
&-& \left. g^Y_\sigma(\sigma(r)) \sigma(r) \right] F_\beta(r) = 0,
\label{diracnr} \\
\frac{d}{dr} F_\beta(r) - \frac{\kappa}{r} F_\beta(r) +
\left[ \epsilon_\beta - g^Y_\omega \omega(r) - t_\beta g^Y_\rho b(r)
\right.
&-& \left. e Q_Y A(r) - M_Y \right. \nn \\
&+& \left. g^Y_\sigma (\sigma(r)) \sigma (r) \right] G_\beta(r) = 0.
\label{diracyr}
\en
Here $G_{\alpha,\beta}(r)/r$ and $F_{\alpha,\beta}(r)/r$ 
are respectively the
radial part of the upper and lower
components of the solution to the Dirac equation for 
the nucleon (hyperon)~\cite{qhd}:
\bge
\psi_{N,Y}({\vec r}) = {i[G_{\alpha,\beta}(r)/r] \Phi_{\kappa m} \choose
-[F_{\alpha,\beta}(r)/r] \Phi_{-\kappa m}} \xi_{t_{\alpha,\beta}},
\label{wave}
\ene
where $\xi_{t_{\alpha,\beta}}$ 
is a two-component isospinor and $\Phi_{\kappa m}$ is
a spin spherical harmonic for the nucleon (hyperon)~\cite{finite2}
($\alpha$ ($\beta$) labelling the quantum numbers of nucleon (hyperon) and 
$\epsilon_{\alpha,\beta}$ being
the energies).  Then, the normalization condition is, 
\bge
\int dr (|G_{\alpha,\beta}(r)|^2 + |F_{\alpha,\beta}(r)|^2) =1. \label{norm}
\ene
As usual, $\kappa$ specifies the angular quantum numbers and $t_\alpha$
($t_\beta$) the eigenvalue of the isospin operator $\tau^N_3/2$ ($I^Y_3$).
The total energy of the hypernuclear system is then given by  
\bg
E_{tot} &=& \sum_\alpha^{occ} (2j_\alpha + 1) \epsilon_\alpha
            + \epsilon_\beta
 - \frac{1}{2} \int d{\vec r} \ [ - g_\sigma C_N(\sigma (r)) \sigma(r)
\rho_s(r) \nn \\
 &+& g_\omega \omega(r) \rho_B(r) + \frac{1}{2} g_\rho b(r) \rho_3(r)
 + eA(r) \rho_p(r) ] \nn \\
 &-& \frac{1}{2} \int d{\vec r} \ [ - g^Y_\sigma C_Y(\sigma (r)) \sigma(r)
\rho^Y_s(r) \nn \\
 &+& g^Y_\omega \omega(r) \rho^Y_B(r) + g^Y_\rho b(r) t_\beta \rho^Y_B(r)
 + eQ_Y A(r) \rho^Y_B(r) ].  \label{ftoten}
\en
%

\section{Spin-orbit potential in the QMC model}\label{sectso}

In this section, we will focus on the spin-orbit potential of 
a hyperon in a hypernucleus.
In order to see the differences between the QMC model 
and the QHD type models, we first discuss the spin-orbit 
potential of the $\Lambda$.

The origin of the spin orbit force for a composite nucleon moving
through scalar and vector fields which vary with position was explained
in detail in Ref.~\cite{finite0} a) -- cf. 
Section 3.2. The situation for the
$\Lambda$ is different in that, in an 
SU(6) quark model, the u and d quarks are coupled 
to spin zero, so that  
the spin of the $\Lambda$ is carried by the s quark. 
As the $\sigma$-meson is viewed here as a convenient parametrization of 
two-pion-exchange and the $\omega$ and $\rho$ are non-strange, it seems 
reasonable to assume that  
the $\sigma$, $\omega$ and $\rho$ mesons couple only to 
the u and d quarks. The direct contributions to the spin-orbit interaction 
from these mesons (derived in Section 3 of Ref.~\cite{finite0} a)) 
then vanish due to the flavor-spin structure.
Thus, the spin-orbit interaction, 
$V^\Lambda_{S.O.}(r) \vec{l}\cdot\vec{s}$, 
at the position $\vec{r}$ of the $\Lambda$ in a hypernucleus 
arises entirely from Thomas precession:  
\begin{equation}
V^\Lambda_{S.O.}(r) \vec{l}\cdot\vec{s} 
= - \frac{1}{2} {\vec{v}_\Lambda} \times 
\frac{d \vec{v_\Lambda}}{dt} \cdot \vec{s}
= - \frac{1}{2 M^{\star 2}_\Lambda (r) r}
\, \left( \frac{d}{dr} [ M^\star_\Lambda (r) 
+ g^\Lambda_\omega \omega(r) ] \right) \vec{l}\cdot\vec{s} ,
\label{so}
\end{equation}
where, $\vec{v}_\Lambda = \vec{p}_\Lambda/M^\star_\Lambda$, 
is the velocity of the $\Lambda$ in the rest frame of the $\Lambda$
hypernucleus, and the acceleration,
$d\vec{v}_\Lambda/dt$, is obtained from the Hamilton equations of motion  
applied to the leading order Hamiltonian of the QMC model~\cite{finite0}.
Because the contributions from the effective mass of the $\Lambda$, 
$M^\star_\Lambda (r)$, 
and the vector potential, $g^\Lambda_\omega \omega(r)$, are 
approximately equal and opposite in sign, 
we quite naturally expect a very 
small spin-orbit interaction for the $\Lambda$ in the hypernucleus. 
Although the spin-orbit splittings for the nucleon calculated 
in QMC are already somewhat smaller~\cite{finite0,finite1} 
than those calculated in QHD~\cite{qhd}, 
we can expect much smaller spin-orbit splittings for the $\Lambda$ 
in QMC, which will be confirmed numerically later. 
In order to include the spin-orbit potential of Eq.~(\ref{so}) correctly,  
we added perturbatively the correction due to the vector potential, 
$ -\frac{2}{2 M^{\star 2}_\Lambda (r) r}
\, \left( \frac{d}{dr} g^\Lambda_\omega \omega(r) \right) 
\vec{l}\cdot\vec{s}$, 
to the single-particle energies obtained with the Dirac 
equation Eq.~(\ref{eqdiracy}), by evaluating it with 
the obtained wave function for 
the $\Lambda$. 
This is necessary because the Dirac 
equation corresponding to Eq.~(\ref{eqdiracy})
leads to a spin-orbit force which does not correspond to the underlying
quark model, namely:
\begin{equation}
V^\Lambda_{S.O.}(r) \vec{l}\cdot\vec{s} 
= - \frac{1}{2 M^{\star 2}_\Lambda (r) r}
\, \left( \frac{d}{dr} [ M^\star_\Lambda (r) 
- g^\Lambda_\omega \omega(r) ] \right) \vec{l}\cdot\vec{s}.
\label{sodirac}
\end{equation}
This correction to the spin-orbit force, which appears naturally in the
QMC model, may also be modelled at the hadronic level of the Dirac equation  
by adding a tensor interaction.

In the QMC model, the general expression for the spin-orbit 
potential felt by the nucleon or hyperon, $j$ 
($j = N, \Lambda, \Sigma, \Xi$), 
may be expressed as

\bg
V^j_{S.O.}(r) \vec{l}\cdot\vec{s} &=& \frac{-1}{2M_j^{\star 2}(r) r}
\left[ \Delta^j_\sigma
  +(G^s_j - 6 F^s_j \mu_s\eta_j(r))\Delta^j_\omega
+ (G^v_j - \frac{6}{5} F^v_j \mu_v\eta_j(r))\Delta^j_\rho \right]
\vec{l}\cdot\vec{s}, \quad
\label{spinorbit}
\en
with
\bg
\Delta^j_\sigma &=& \frac{d}{dr} M^{\star}_j(r), \quad 
\Delta^j_\omega = \frac{d}{dr} (\frac{1}{3}) g_\omega \omega(r) 
= \Delta^q_\omega = \frac{d}{dr} g^q_\omega \omega(r), \quad
\Delta^j_\rho =\frac{d}{dr} g_\rho b(r), \label{delta} \\     
\nn \\
G^s_j &=& <j| \sum_{i=u,d} 1(i) |j>, \qquad
G^v_j = <j| \sum_{i=u,d} \frac{1}{2} \tau_3 (i) |j>, \label{gsv} \\
F^s_j &=& \frac{<j| \sum_{i=u,d} \frac{1}{2} \vec{\sigma}(i) |j>}
{<j| \frac{1}{2} \vec{\sigma}^j |j>}, \qquad
F^v_j = \frac{<j| \sum_{i=u,d} \frac{1}{2} \vec{\sigma}(i) 
\frac{1}{2} \tau_3(i) |j>}
{<j| \frac{1}{2} \vec{\sigma}^j |j>}, \label{fsv} \\ 
\nn \\
\mu_s &=& \frac{1}{3} M_N I_0 = \frac{1}{5} \mu_v, \qquad 
\eta_j(r) = \frac{I_j^\star M_j^\star (r)}{I_0 M_N},  \label{mueta} \\
I_0 &=& \frac{R_N}{3} \frac{4 \Omega_N + 2 m_{u,d}R_N - 3}
{2 \Omega_N (\Omega_N - 1) + m_{u,d} R_N}, \label{int0}  \\
I_j^\star &=& \frac{R_j^\star}{3} \frac{4 \Omega_j^\star (\sigma) +
2 m^\star_{u,d}(\sigma) R_j^\star - 3}
{2 \Omega_j^\star (\sigma) (\Omega_j^\star (\sigma) - 1) 
+ m^\star_{u,d}(\sigma) R_j^\star}. \label{intj}
\en
Here $\vec{r}$ is the position of the baryon $j$ in the 
hypernucleus (nucleus), and 
the terms proportional to $\mu_s$ and $\mu_v$ are the anomalous 
contributions from the u and/or d quarks  
due to the finite size of the hyperon (nucleon). 
They are related to the magnetic moments 
of the proton, $\mu_p$, and the neutron, $\mu_n$, as 
$\mu_s = \mu_p + \mu_n$ and $\mu_v = \mu_p - \mu_n$, with the experimental 
values, $\mu_p = 2.79$ and $\mu_n = -1.91$ (in nuclear magnetons). 
However, these $\mu_s$ and $\mu_v$, together with the quantities 
$I_0$ and $I_j^\star$ of Eqs.~(\ref{int0}) and (\ref{intj}), 
must be, and will be calculated self-consistently in the model. 
Note that the in-medium bag radius, $R_j^\star$, and 
the lowest bag eigenenergy for the u (d) quark, $\Omega_j^\star$ 
in units of $1/R_j^\star$,  
depend on $j$ $(j = N,\Lambda,\Sigma,\Xi)$,   
because the bag radius is generally 
different for each iso-multiplet of the octet baryons in the medium. 
The explicit expressions for the spin-orbit potentials 
for the octet baryons in the QMC model are given by:
\bg
V^p_{S.O.}(r) &=& \frac{-1}{2M_N^{\star 2}(r) r}
\left[ \Delta^N_\sigma
  + 3 (1 - 2\mu_s\eta_N(r))\Delta^N_\omega
+ \frac{1}{2}(1 - 2\mu_v\eta_N(r))\Delta_\rho \right], 
\label{sop} \\
V^n_{S.O.}(r) &=& \frac{-1}{2M_N^{\star 2}(r) r}
\left[ \Delta^N_\sigma
  + 3 (1 - 2\mu_s\eta_N(r))\Delta^N_\omega
- \frac{1}{2}(1 - 2\mu_v\eta_N(r))\Delta_\rho \right],
\label{son} \\
V^\Lambda_{S.O.}(r) &=& \frac{-1}{2M_\Lambda^{\star 2}(r) r}
\left[ \Delta^\Lambda_\sigma
  + 2 \Delta^\Lambda_\omega \right],
\label{sola} \\
V^{\Sigma^+}_{S.O.}(r) &=& \frac{-1}{2M_\Sigma^{\star 2}(r) r}
\left[ \Delta^\Sigma_\sigma
  + 2 (1 - 4\mu_s\eta_\Sigma(r))\Delta^\Sigma_\omega
+ (1 - \frac{4}{5}\mu_v\eta_\Sigma(r))\Delta_\rho \right],
\label{sos+} \\
V^{\Sigma^0}_{S.O.}(r) &=& \frac{-1}{2M_\Sigma^{\star 2}(r) r}
\left[ \Delta^\Sigma_\sigma
  + 2 (1 - 4\mu_s\eta_\Sigma(r))\Delta^\Sigma_\omega \right], 
\label{sos0} \\
V^{\Sigma^-}_{S.O.}(r) &=& \frac{-1}{2M_\Sigma^{\star 2}(r) r}
\left[ \Delta^\Sigma_\sigma
  + 2 (1 - 4\mu_s\eta_\Sigma(r))\Delta^\Sigma_\omega
- (1 - \frac{4}{5}\mu_v\eta_\Sigma(r))\Delta_\rho \right],
\label{sos-} \\
V^{\Xi^0}_{S.O.}(r) &=& \frac{-1}{2M_\Xi^{\star 2}(r) r}
\left[ \Delta^\Xi_\sigma
  +(1 + 2\mu_s\eta_\Xi(r))\Delta^\Xi_\omega
+ \frac{1}{2}(1 + \frac{2}{5}\mu_v\eta_\Xi(r))\Delta_\rho \right],
\label{sox0} \\
V^{\Xi^-}_{S.O.}(r) &=& \frac{-1}{2M_\Xi^{\star 2}(r) r}
\left[ \Delta^\Xi_\sigma
  +(1 + 2\mu_s\eta_\Xi(r))\Delta^\Xi_\omega
- \frac{1}{2}(1 + \frac{2}{5}\mu_v\eta_\Xi(r))\Delta_\rho \right].
\label{sox-} 
\en

In the next Section, we will show the
spin-orbit potentials for $\Lambda, \Sigma^0$ and $\Xi^0$ in
$^{41}_Y$Ca and $^{209}_Y$Pb hypernuclei, which are evaluated 
self-consistently in the QMC model.  
These will be shown for the $1p_{3/2}$
hyperon configuration, for which the change in 
the self-consistently calculated meson mean fields
is expected to be the largest among the states which have 
non-zero spin-orbit interactions.

\section{Numerical results within the QMC model alone}\label{numeri}

In this section, we will discuss some of the results calculated in the 
QMC model alone, in order to extract the modifications necessary  
to achieve a more realistic calculation.

First, we need to specify the parameters used in the calculation.
The parameters at the hadronic level, which are already 
fixed by the study of
infinite nuclear matter and finite nuclei~\cite{finite1},
are shown in Table~\ref{hparamt}.
%
\begin{table}[htbp]
\begin{center}
\caption{Parameters at the hadronic
level (for finite nuclei)~\protect\cite{finite1}.}
\label{hparamt}
\begin{tabular}[t]{c|cc}
\hline
field &mass (MeV) &$g^2/4\pi\, (e^2/4\pi)$\\
\hline
$\sigma$ &418 &3.12\\
$\omega$ &783 &5.31\\
$\rho$   &770 &6.93\\
$A$      &0   &1/137.036\\
\end{tabular}
\end{center}
\end{table}
%
%
Concerning the parameters for the $\sigma$ meson, we note that 
the properties of nuclear matter only fix the ratio, ($g_\sigma/m_\sigma$), 
with a chosen value, $m_\sigma = 550$ MeV. 
Keeping this ratio to be a constant, 
the value $m_\sigma = 418$ MeV for finite nuclei (hypernuclei) 
is obtained by fitting the r.m.s. charge radius of $^{40}$Ca  
to the experimental value, $r_{{\rm ch}}(^{40}$Ca) = 3.48 fm~\cite{finite1}.

Next, we show the calculated single-particle 
energies for $^{209}_Y$Pb hypernuclei 
in Table~\ref{speqmc}. 
%
\begin{table}[htbp]
\begin{center}
\caption{Single-particle energies (in MeV)
for $^{209}_Y$Pb hypernuclei for $Y = \Lambda, \Sigma$ and $\Xi$. 
$^{209}_\Lambda$Pb$^\star$ denotes the results 
calculated with using the scaled coupling constant, 
$0.93 \times g_\sigma^\Lambda (\sigma = 0)$, which reproduces the 
empirical single-particle energy for the $1s_{1/2}$ in $^{41}_\Lambda$Ca, 
-20.0 MeV~\protect\cite{chr}.
The experimental data for
$^{208}_\Lambda$Pb are taken from Ref.~\protect\cite{aji}.
Spin-orbit splittings are not well determined by the experiments.}
\label{speqmc}
\begin{tabular}[t]{c|cccccccc}
\hline \hline
&$^{208}_\Lambda$Pb (Expt.)
&$^{209}_\Lambda$Pb$^\star$ &$^{209}_\Lambda$Pb &$^{209}_{\Sigma^-}$Pb
&$^{209}_{\Sigma^0}$Pb &$^{209}_{\Sigma^+}$Pb
&$^{209}_{\Xi^-}$Pb    &$^{209}_{\Xi^0}$Pb\\ 
\hline \hline
$1s_{1/2}$&-27.0        &-27.4 &-35.9 &-46.4 &-35.8 &-25.8 &-33.5 &-23.6\\
$1p_{3/2}$&             &-23.8 &-31.9 &-41.7 &-32.0 &-22.6 &-29.7 &-20.5\\
$1p_{1/2}$&-22.0 $(1p)$ &-23.8 &-31.9 &-41.6 &-31.9 &-22.5 &-29.7 &-20.5\\
$1d_{5/2}$&             &-19.5 &-27.1 &-36.8 &-27.6 &-18.6 &-25.6 &-16.8\\
$2s_{1/2}$&             &-17.9 &-25.4 &-36.2 &-25.9 &-15.9 &-25.1 &-15.7\\
$1d_{3/2}$&-17.0 $(1d)$ &-19.5 &-27.1 &-36.6 &-27.4 &-18.3 &-25.6 &-16.8\\
$1f_{7/2}$&             &-14.7 &-21.8 &-31.5 &-22.6 &-13.9 &-21.2 &-12.7\\
$2p_{3/2}$&             &-12.6 &-19.4 &-29.7 &-20.3 &-10.9 &-20.4 &-11.3\\
$1f_{5/2}$&-12.0 $(1f)$ &-14.6 &-21.7 &-31.1 &-22.3 &-13.5 &-21.3 &-12.8\\
$2p_{1/2}$&             &-12.6 &-19.4 &-29.5 &-20.1 &-10.7 &-20.4 &-11.4\\
$1g_{9/2}$&             &-9.5  &-16.0 &-25.8 &-17.2 &-8.6  &-16.7 &-8.3 \\
$1g_{7/2}$&-7.0  $(1g)$ &-9.4  &-15.9 &-25.3 &-16.6 &-8.1  &-16.8 &-8.4 \\
$1h_{11/2}$&            &-4.0  &-9.8  &-19.9 &-11.5 &-3.0  &-12.1 &-3.7 \\
$2d_{5/2}$&             &-7.1  &-13.2 &-23.6 &-14.4 &-5.2  &-15.8 &-6.7 \\
$2d_{3/2}$&             &-7.1  &-13.2 &-23.4 &-14.1 &-4.9  &-15.9 &-6.7 \\
$1h_{9/2}$&             &-3.9  &-9.7  &-19.2 &-10.7 &-2.2  &-12.2 &-3.8 \\
$3s_{1/2}$&             &-6.2  &-12.1 &-23.5 &-13.2 &-3.1  &-15.7 &-5.9 \\
$2f_{7/2}$&             &-1.8  &-6.9  &-17.7 &-8.4  &---   &-11.6 &-2.0 \\
$3p_{3/2}$&             &-1.1  &-5.6  &-16.9 &-7.0  &---   &-7.4  &-1.4 \\
$2f_{5/2}$&             &-1.7  &-6.8  &-17.3 &-8.0  &---   &-11.7 &-2.0 \\
$3p_{1/2}$&             &-1.1  &-5.6  &-16.8 &-6.8  &---   &-7.4  &-1.4 \\
$1i_{13/2}$&            &---   &-3.4  &-13.9 &-5.4  &---   &-7.6  &---  \\
\end{tabular}
\end{center}
\end{table}
%
%
In principle, the existence of the hyperon outside of the nuclear core 
breaks spherical symmetry, and one should include this in a truly rigorous
treatment. We neglected this effect, since it is expected to be 
of little importance for spectroscopic calculations~\cite{fur,coh}. 
However, we included the response of the nuclear core 
arising from the self-consistent 
calculation, which is significant for a description of the baryon 
currents and magnetic moments, and a purely relativistic 
effect~\cite{coh,coh2}. Thus, we should always specify the 
state of the hyperon in which the calculation was performed.

Concerning the single-particle energy levels for 
$\Lambda$ hypernuclei, although a direct comparison 
with the data is not precise due to the
different configurations from those observed in the experiments, 
the calculated results seem to significantly 
overestimate the binding. In order to make an estimate of 
the difference due to this different configuration, we calculated 
the following quantity. By removing one $1p_{3/2}$ neutron in 
$^{16}$O, and putting a $\Lambda$ as experimentally observed, 
we calculate in the same way as for $^{17}_\Lambda$O - which 
means the nuclear core is still treated as spherical, although 
the core nucleus is deformed, and a rearrangement of the shell structure
is expected to happen.
In this case the calculated energy for the $1s_{1/2}$ $\Lambda$ is -19.9 MeV, 
to be compared with the value -20.5 MeV of the present treatment.

At first look, the spin-orbit splittings for the $\Lambda$ 
single-particle energies for $^{209}_\Lambda$Pb  
almost vanish, as explained in Section~\ref{sectso}.
This is a very successful consequence of the explicit quark 
structure of the $\Lambda$ considered in the 
SU(6) quark model.

We repeated the calculation using the scaled coupling constant,
$0.93 \times g_\sigma^\Lambda (\sigma =0)$,  
which reproduces the
empirical single-particle energy for the $1s_{1/2}$ in
$^{41}_\Lambda$Ca, -20.0 MeV~\cite{chr}.
The results obtained using this scaled coupling constant
are denoted by $^{209}_\Lambda$Pb$^\star$ in Table~\ref{speqmc}.
Then, one can easily understand that the QMC model does not
require a large SU(3)
breaking effect (only 7 \%) to reproduce the
empirical single-particle energies.
We should mention that a similar calculation was carried out by varying the 
coupling constant, $g^\Lambda_\omega$. In this case, the coupling  
constant obtained was, $1.10 \times g_\omega^\Lambda$, to reproduce 
the empirical single-particle energy for the $1s_{1/2}$ in
$^{41}_\Lambda$Ca, -20.0 MeV~\cite{chr}. 
The calculated results with this scaled coupling constant,
$1.10 \times g_\omega^\Lambda$, are 
almost identical to those calculated with the coupling constant, 
$0.93 \times g_\sigma^\Lambda$. This shows again that the QMC model does 
not require large SU(3) symmetry breaking to reproduce the 
empirical data.

Although the present treatment is based on the underlying quark structure of 
the nucleon and hyperon, the overestimates discussed above 
may be ascribed to 
the lack of an explicit inclusion of the Pauli principle 
at the quark level among the u and d quarks in the core nucleons and 
the $\Lambda$. If one wants to reproduce the experimental 
single-particle energies, one needs to include the Pauli principle 
at the quark level which gives some repulsion. 

For $\Sigma$ and $\Xi$ hypernuclei, the situation is even more 
complicated. We need to 
take into account the effect of channel coupling, 
$\Sigma N - \Lambda N$ and $\Xi N - \Lambda \Lambda$, as well as 
the Pauli blocking effect at the quark level. In particular, the Pauli 
blocking effect is significant in the open, coupled $\Lambda N$ 
channel~\cite{dov,joh,dovy,yyam,mor,hau,ose,afn,afn2,aka,yam}, whereas 
there are no such open channels for two nucleons in the 
core of a hypernucleus. 
The results shown in Table~\ref{speqmc} (and also Table~\ref{rmst})   
are obtained without including either these channel coupling effects, 
or the Pauli blocking effect at the quark level. 
These effects will be considered in specific ways at the 
hadronic level in Section 5.

In Table~\ref{rmst}, we list the calculated binding energy per baryon 
(including hyperon), $-E/A$, 
r.m.s. charge radius, $r_{ch}$, and r.m.s. radii of the the hyperon $Y$ and 
the neutron and proton distributions ($r_Y$, $r_n$ and $r_p$, 
respectively), for the $1s_{1/2}$ and $1p_{3/2}$
hyperon configurations. The r.m.s. charge 
radius is calculated by convolution
with a proton form factor~\cite{finite1}.
For comparison, we also give these quantities without 
a hyperon - i.e., for normal finite nuclei. The differences in values 
for finite nuclei and hypernuclei listed in Table~\ref{rmst} 
reflect the effects of the hyperon, 
through the self-consistency procedure. One can easily see that the effects of 
the hyperon become weaker as the atomic number becomes larger,  
and as the hyperon binding energies become smaller.
We should comment on the small change in the 
r.m.s. radius of the protons, $r_p$, 
in Table~\ref{rmst}. This appears to be a very subtle effect, related to 
the fact that the proton levels in the core are  
relatively shallow and closer to each other compared with those 
calculated in QHD~\cite{finite1}. Thus, the proton 
wave functions in the QMC model are already pushed outside 
compared to those of QHD. We therefore expect then 
to be much more sensitive 
to the change in the mean field potentials 
and especially to the relative change of the Coulomb potential. 
We believe that this is the reason why $r_p$ increases. 
For example, when the Coulomb potential in the core of the hypernucleus 
was artificially switched off 
(but the potential from the $\rho$ meson was kept),
both $r_n$ and $r_p$ decreased for the lowest, $1s1/2$, hypernucleus state.

%
%
\begin{table}[htbp]
\begin{center}
\caption{Binding energy per baryon, $-E/A$ (in MeV), r.m.s. charge radius,
$r_{ch}$, and r.m.s. radii of the hyperon, $r_Y$,
neutron, $r_n$, and proton, $r_p$ (in fm) for $^{41}_Y$Ca 
and $^{209}_Y$Pb. $^{41}_\Lambda$Ca$^\star$ and $^{209}_\Lambda$Pb$^\star$
are the results calculated with using the scaled coupling constant,
$0.93 \times g_\sigma^\Lambda (\sigma = 0)$, which reproduces the
empirical single-particle energy for the $1s_{1/2}$ in $^{41}_\Lambda$Ca,
-20.0 MeV~\protect\cite{chr}. 
Double asterisks, $^{**}$, indicates 
the value used for fitting.}
\label{rmst}
\begin{tabular}[t]{ccccccc}
\\
\hline \hline
hypernuclei &hyperon state& $-E/A$ & $r_{ch}$ & $r_Y$ & $r_n$ & $r_p$ \\
\hline \hline
$^{41}_\Lambda$Ca$^\star$    &$1s_{1/2}$ &7.66&3.51&2.76&3.31&3.42\\
$^{41}_\Lambda$Ca    &$1s_{1/2}$ &7.83&3.52&2.55&3.30&3.42\\
$^{41}_{\Sigma^+}$Ca &$1s_{1/2}$ &7.65&3.52&2.69&3.31&3.42\\
$^{41}_{\Sigma^0}$Ca &$1s_{1/2}$ &7.83&3.52&2.52&3.30&3.42\\
$^{41}_{\Sigma^-}$Ca &$1s_{1/2}$ &7.97&3.52&2.31&3.30&3.42\\
$^{41}_{\Xi^0}$Ca    &$1s_{1/2}$ &7.43&3.50&2.80&3.32&3.40\\
$^{41}_{\Xi^-}$Ca    &$1s_{1/2}$ &7.54&3.50&2.62&3.32&3.40\\
\hline
$^{41}_\Lambda$Ca$^\star$    &$1p_{3/2}$ &7.48&3.50&3.43&3.32&3.41\\
$^{41}_\Lambda$Ca    &$1p_{3/2}$ &7.57&3.51&3.17&3.32&3.41\\
$^{41}_{\Sigma^+}$Ca &$1p_{3/2}$ &7.47&3.51&3.24&3.32&3.41\\
$^{41}_{\Sigma^0}$Ca &$1p_{3/2}$ &7.59&3.51&3.13&3.32&3.41\\
$^{41}_{\Sigma^-}$Ca &$1p_{3/2}$ &7.70&3.51&3.02&3.31&3.41\\
$^{41}_{\Xi^0}$Ca    &$1p_{3/2}$ &7.30&3.49&3.58&3.33&3.39\\
$^{41}_{\Xi^-}$Ca    &$1p_{3/2}$ &7.40&3.49&3.36&3.32&3.39\\
\hline
$^{40}$Ca            & ---       &7.36&3.48$^{**}$&--- &3.33&3.38\\
\hline \hline
$^{209}_\Lambda$Pb$^\star$    &$1s_{1/2}$ &7.36&5.49&3.97&5.67&5.43\\
$^{209}_\Lambda$Pb    &$1s_{1/2}$ &7.39&5.50&3.79&5.67&5.43\\
$^{209}_{\Sigma^+}$Pb &$1s_{1/2}$ &7.37&5.49&4.06&5.67&5.43\\
$^{209}_{\Sigma^0}$Pb &$1s_{1/2}$ &7.39&5.50&3.77&5.67&5.43\\
$^{209}_{\Sigma^-}$Pb &$1s_{1/2}$ &7.41&5.50&3.42&5.67&5.43\\
$^{209}_{\Xi^0}$Pb    &$1s_{1/2}$ &7.32&5.49&3.82&5.68&5.43\\
$^{209}_{\Xi^-}$Pb    &$1s_{1/2}$ &7.34&5.49&3.48&5.68&5.43\\
\hline
$^{209}_\Lambda$Pb$^\star$    &$1p_{3/2}$ &7.34&5.49&4.70&5.68&5.43\\
$^{209}_\Lambda$Pb    &$1p_{3/2}$ &7.37&5.49&4.50&5.67&5.43\\
$^{209}_{\Sigma^+}$Pb &$1p_{3/2}$ &7.34&5.49&4.65&5.68&5.43\\
$^{209}_{\Sigma^0}$Pb &$1p_{3/2}$ &7.37&5.49&4.46&5.67&5.43\\
$^{209}_{\Sigma^-}$Pb &$1p_{3/2}$ &7.39&5.49&4.28&5.67&5.43\\
$^{209}_{\Xi^0}$Pb    &$1p_{3/2}$ &7.30&5.49&4.57&5.68&5.43\\
$^{209}_{\Xi^-}$Pb    &$1p_{3/2}$ &7.32&5.49&4.35&5.68&5.43\\
\hline
$^{208}$Pb            & ---       &7.25&5.49&--- &5.68&5.43\\
\end{tabular}
\end{center}
\end{table}
%
%

\begin{figure}[hbt]
\begin{center}
\epsfig{file=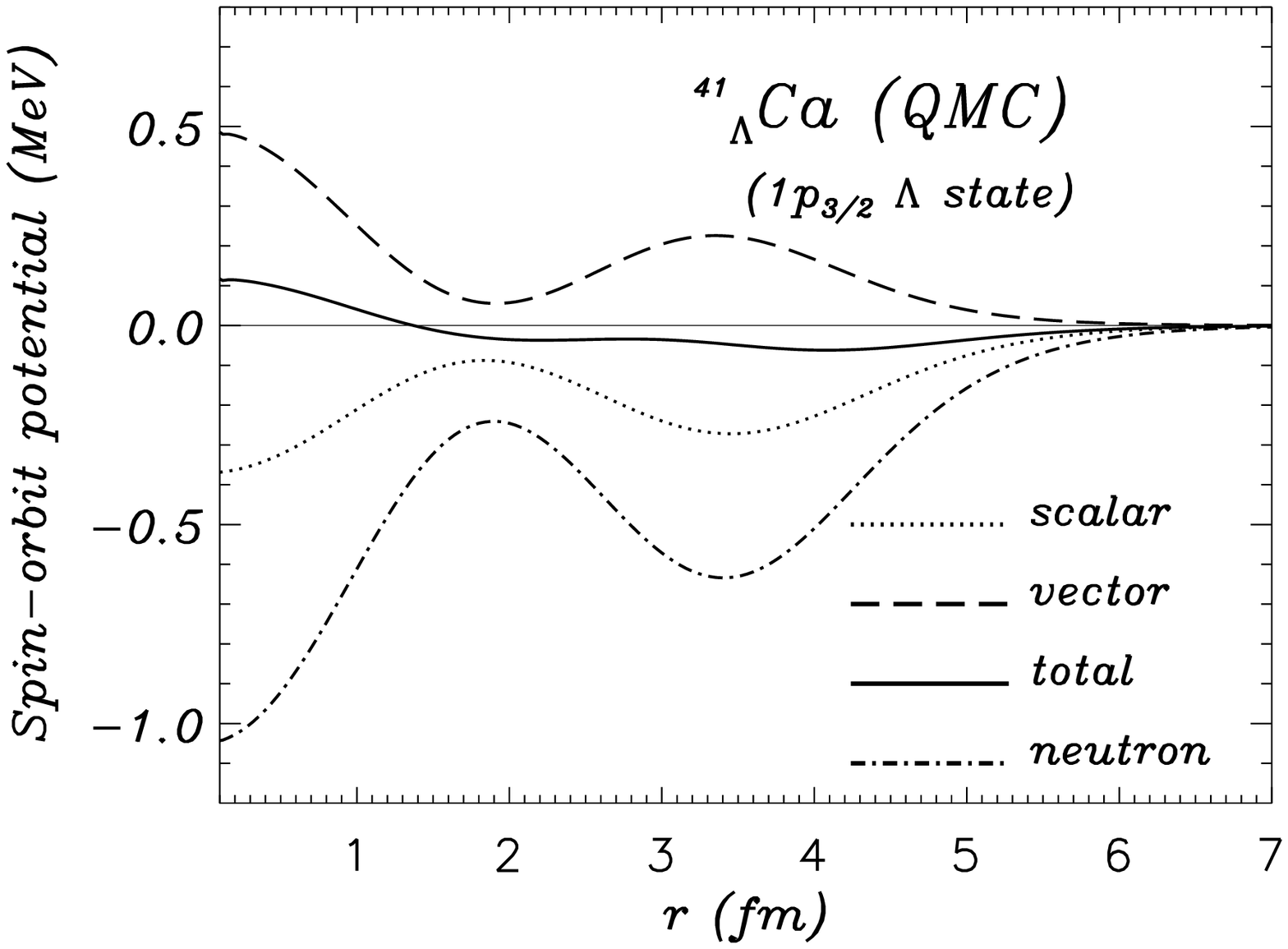,height=7cm}
\epsfig{file=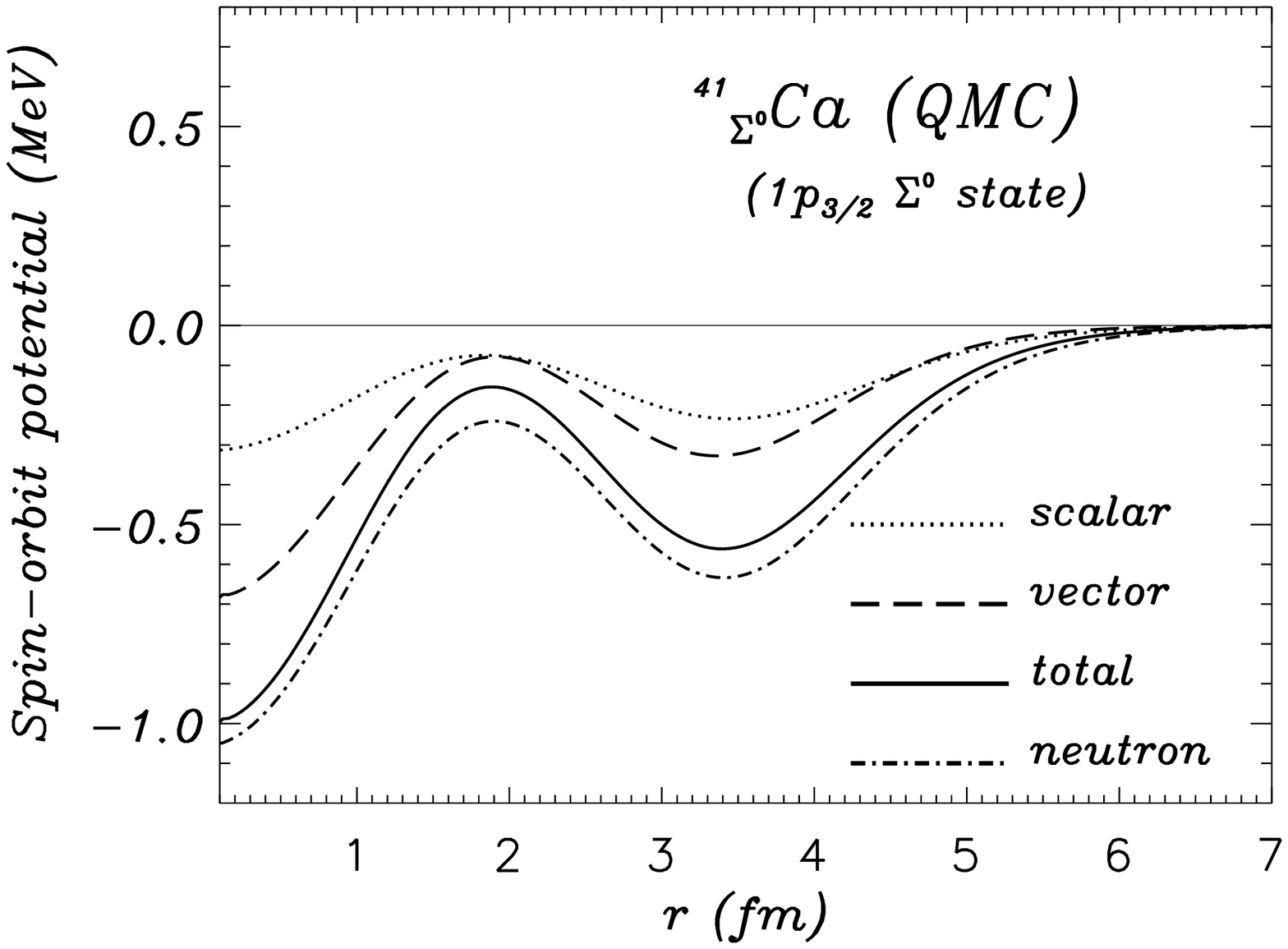,height=7cm}
\epsfig{file=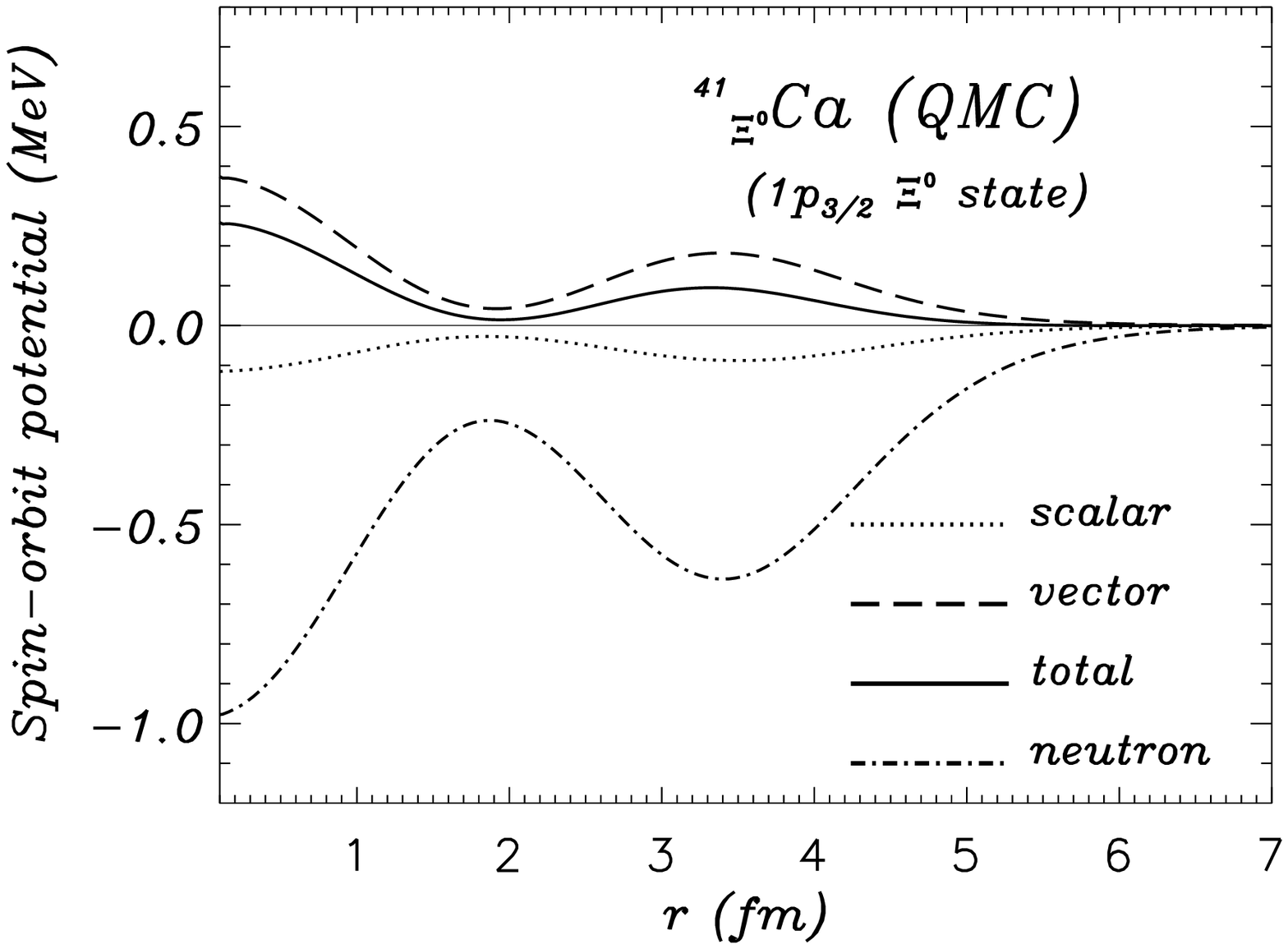,height=7cm}
\caption{Spin-orbit potentials, $V^Y_{S.O.}(r)$,  
calculated in the QMC model alone 
for $Y = \Lambda$, $\Sigma^0$ and $\Xi^0$ in 
$^{41}_Y$Ca hypernuclei for the $1p_{3/2}$ hyperon state.}
\label{soca}
\end{center}
\end{figure}

\begin{figure}[hbt]
\begin{center}
\epsfig{file=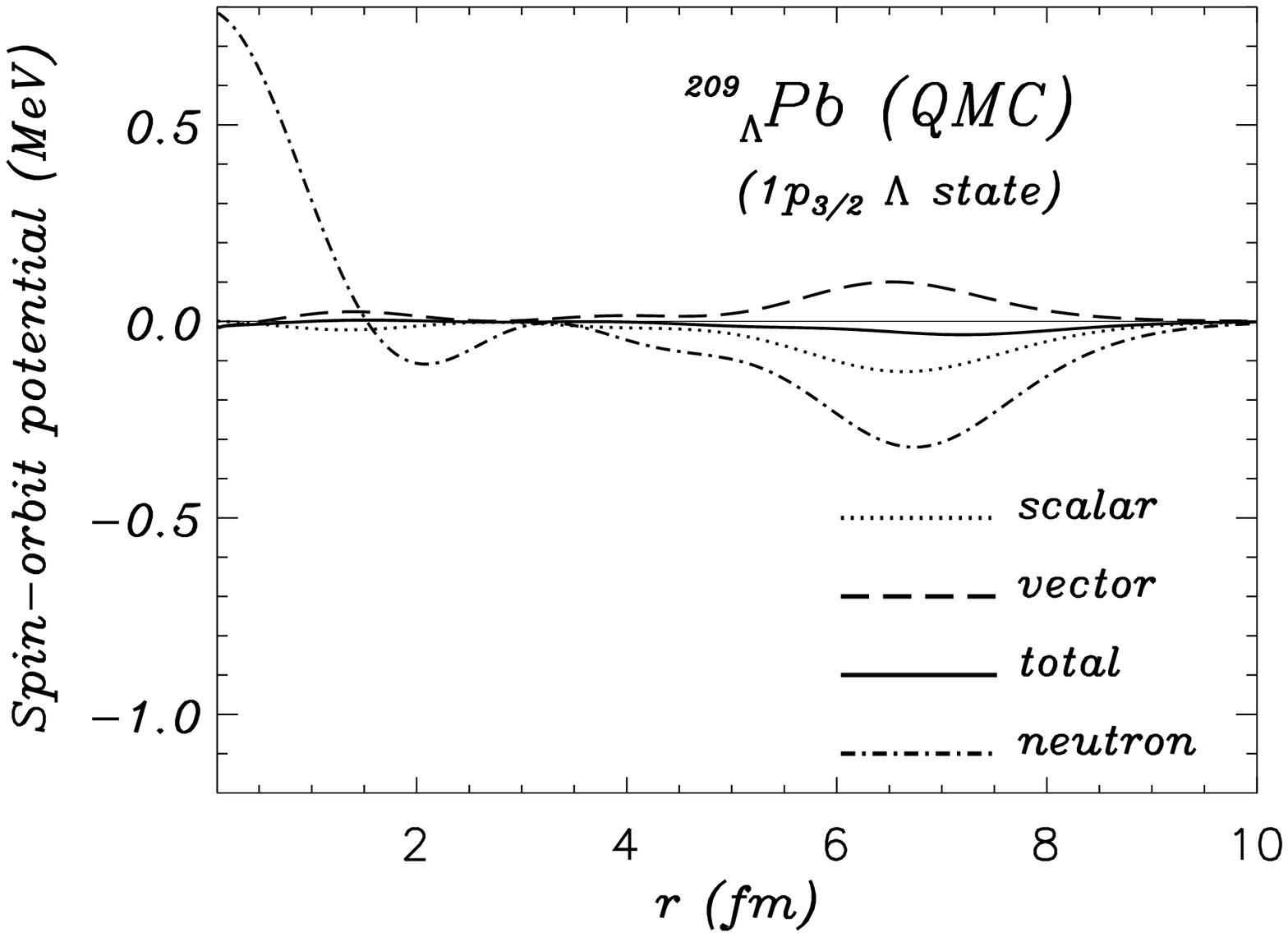,height=7cm}
\epsfig{file=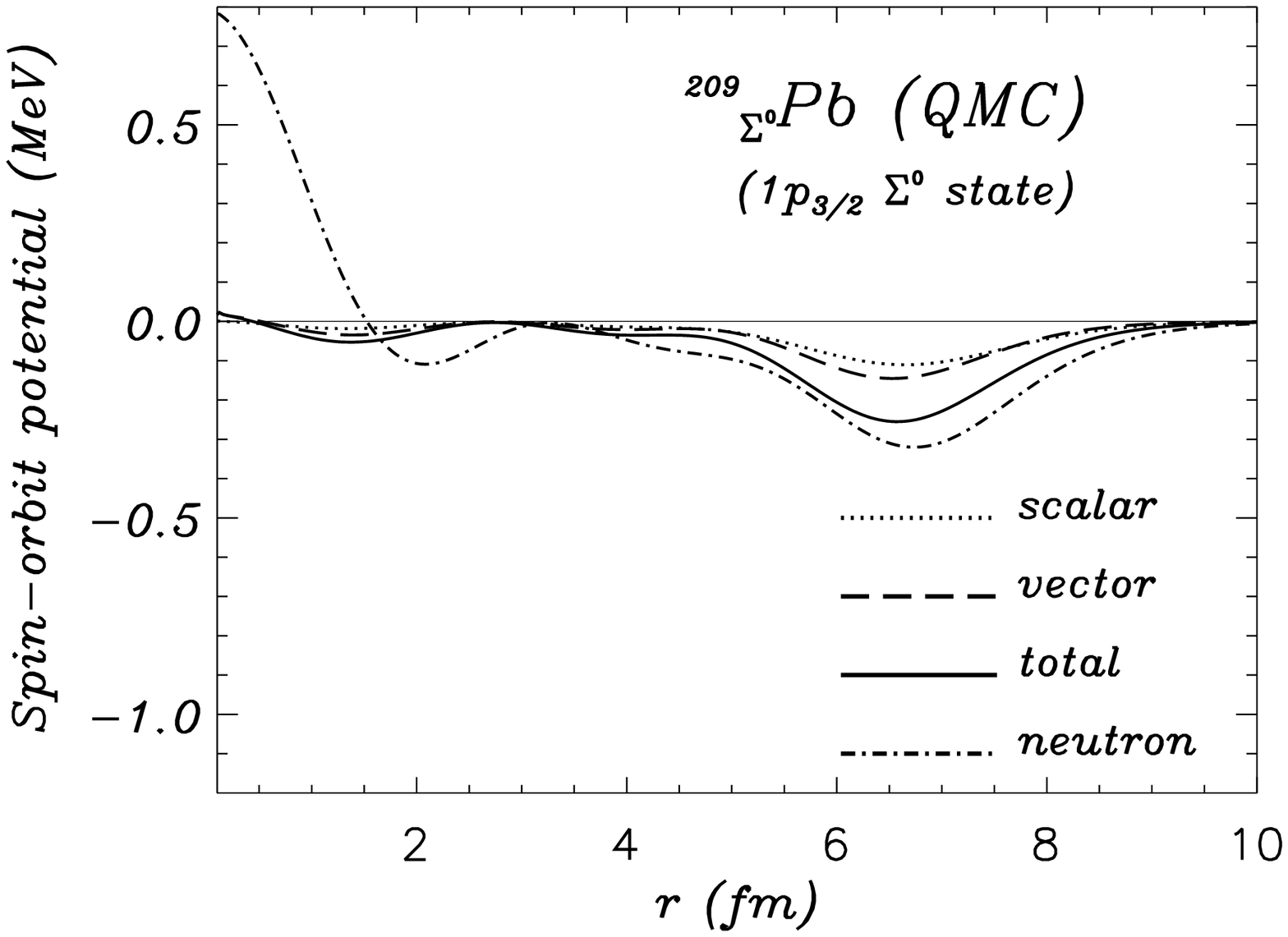,height=7cm}
\epsfig{file=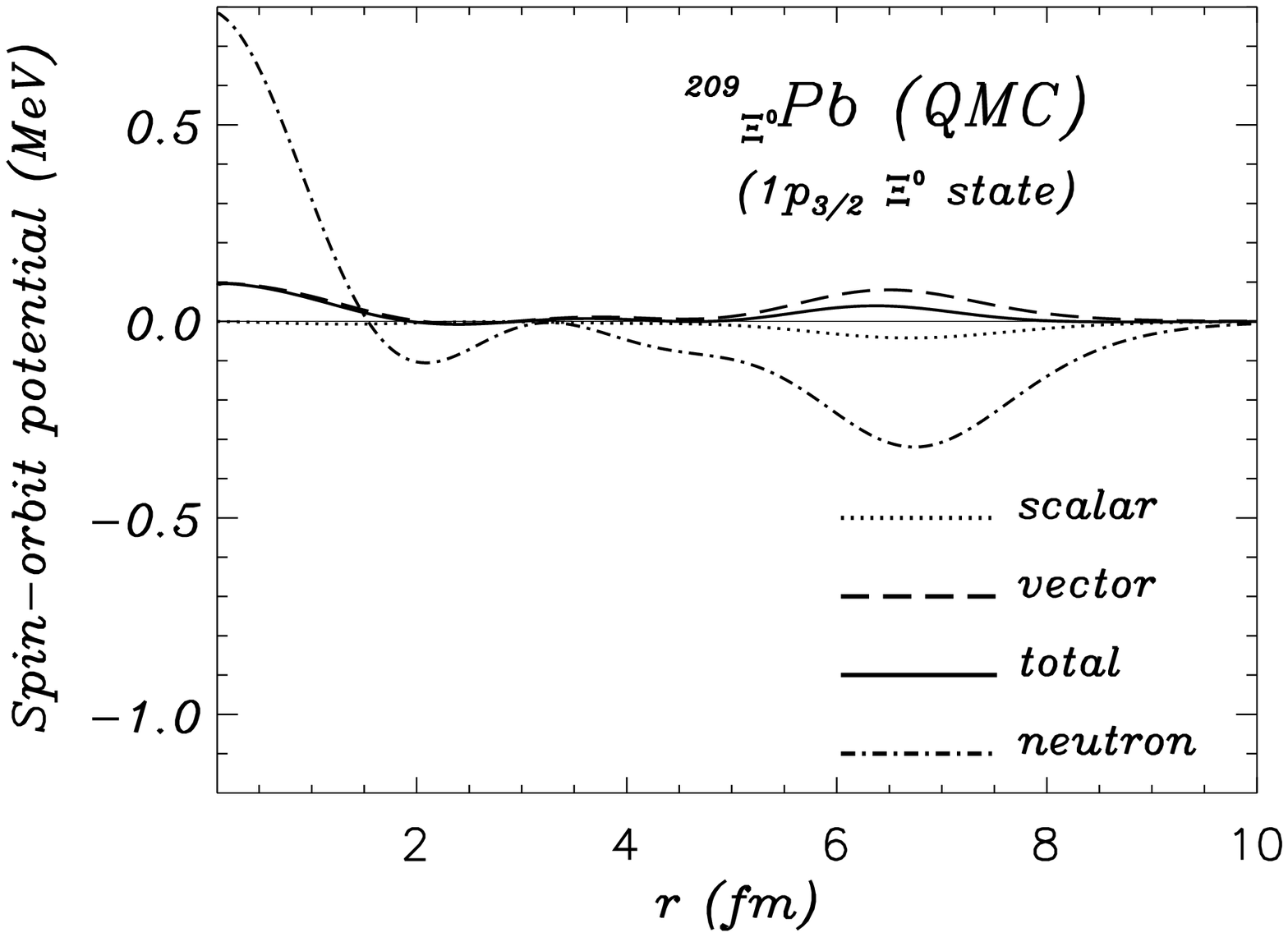,height=7cm}
\caption{Spin-orbit potentials, $V^Y_{S.O.}(r)$, 
calculated in the QMC model alone 
for $Y = \Lambda$, $\Sigma^0$ and $\Xi^0$ in 
$^{209}_Y$Pb hypernuclei for the $1p_{3/2}$ hyperon state.}
\label{sopb}
\end{center}
\end{figure}

Next we show in Figs.~\ref{soca} and \ref{sopb} 
the spin-orbit potentials for the hyperons,  
$\Lambda$, $\Sigma^0$ and $\Xi^0$, in $^{41}_Y$Ca and $^{209}_Y$Pb
hypernuclei for the $1p_{3/2}$ hyperon configuration. 
For comparison, we also show the spin-orbit potential for the neutron 
calculated self-consistently by Eq.~(\ref{son}),  
with the same configuration as the hyperon.

%
%
%
%

%
%
%
%

It is interesting to focus on some particular single-particle energy 
levels, e.g., $1g_{9/2}$ and $1g_{7/2}$ in $^{209}_\Lambda$Pb, 
$^{209}_{\Sigma^0}$Pb and $^{209}_{\Xi^0}$Pb. Comparing  
these single-particle energies with 
the plotted spin-orbit potentials in 
Fig.~\ref{sopb}, one can understand the origin of 
the reverse in magnitudes for the single-particle energies between 
the $^{209}_{\Sigma^0}$Pb ($^{209}_\Lambda$Pb) and 
$^{209}_{\Xi^0}$Pb, and why the spin-orbit splittings 
for $^{209}_{\Sigma^0}$Pb are the most pronounced ones among them. 
Although the spin-orbit potentials shown in 
Fig.~\ref{sopb} are for the $1p_{3/2}$ hyperon configuration, 
the main features 
also hold for these energy levels.

As for the spin-orbit potentials for the $\Sigma$ in hypernuclei, 
the QMC results  
show a slight inequality $|V^\Sigma_{S.O.}| < |V^N_{S.O.}|$, 
which is indicated in Fig.~\ref{soca}. 
This contrasts with the result of Pirner~\cite{pir},  
$|V^\Sigma_{S.O.}| = \frac{4}{3} |V^N_{S.O.}|$, 
which did not include the contribution from the Thomas precession 
($G^s_j$ term in Eq.~(\ref{spinorbit})), but is 
rather similar to the results of Brockmann~\cite{bro2} and 
Dover, Millener and Gal~\cite{dovy}.
(Note that the ratio $\frac{4}{3}$ of Ref.~\cite{pir} does agree 
with the ratio $F^s_\Sigma/F^s_N$ alone.) 
Finally, we should also note that the vector potential 
adopted by Pirner~\cite{pir} 
is based on the combined quark-gluon exchange. 
Thus, the corresponding spin-orbit potential of ours,
$F^s_j$ term in Eq.~(\ref{spinorbit}), has a different origin.

%
\section{Effects of Pauli blocking and channel coupling}

As discussed in Section~\ref{numeri}, when we insert a hyperon into an 
orbital with the same quantum numbers as one already occupied 
by nucleons we need to include 
the effects of Pauli blocking at the quark level. 
In the case of the $\Sigma$-hyperon there is an additional correction 
associated with the coupling to the open $\Lambda N$ channel.
These effects will be considered in specific ways at the hadronic level
in this Section. A more consistent treatment 
at the quark level is beyond the scope of the present study.

%
\subsection{The Pauli blocking effect}\label{pauli}
Here we consider the effect of the Pauli blocking at the quark level 
in a specific way at the hadronic level.
This effect is significant in the open, coupled channel, 
particularly in the $\Sigma N - \Lambda N$ channel in the present case, 
where such phenomenon simply does not occur for two nucleons 
in the core nucleus. Thus, the Pauli blocking effect 
we include in the present study should be regarded  
in order to ensure the correct channel coupling effect. 

It seems natural to assume that this effect works 
repulsively in a way that the strength is proportional  
to the u and/or d quark baryonic (number) density of the core nucleons.
Then one can expect that   
the u and/or d quarks in the hyperon 
feel a stronger repulsion at the position where the 
baryon density is large. As a consequence, 
the wave function of the hyperon (quark) 
will be suppressed in this region.
We assume here that the Pauli blocking effect is simply proportional to the 
baryonic density, although one could 
consider a more complicated density dependence.
Then, the Dirac equation for the hyperon Y, Eq.~(\ref{eqdiracy}),  
may be modified by 
\bge
[i\gamma \cdot \partial - M^{\star}_Y(\sigma)-
(\, \lambda_Y \rho_B(\vec{r}) + g^Y_\omega \omega(\vec{r}) 
+ g_\rho I^Y_3 b(\vec{r})
+ e Q_Y A(\vec{r}) \,)
\gamma_0 ] \psi_Y(\vec{r}) = 0, \label{mdiracy}
\ene
where, $\rho_B(\vec{r})$ is the baryonic density at the position $\vec{r}$ 
in the hypernucleus due to the core nucleons, and $\lambda_Y$ is a 
constant to be determined empirically. (We note that the Pauli
effect associated with the light quarks in the $\Lambda$ should also 
lead to some repulsion for the nucleons. As this will be an ${\cal O}
(\frac{1}{A})$ effect on the nucleon levels we have neglected this
reaction back on the nucleons in the present work.)
In the present treatment, we chose this 
constant $\lambda_Y, Y=\Lambda$, in order to reproduce the empirical single 
particle energy for the 
$1s_{1/2}$ in $^{209}_\Lambda$Pb,  
-27.0 MeV~\cite{aji}.

One might also imagine that this fitted value includes the attractive 
$\Lambda N \rightarrow \Sigma N$
channel coupling effect for the $\Lambda$ single-particle enegies, 
because the value fitted is the experimentaly observed one.
However, for $\Sigma$ hypernuclei, the repulsive 
$\Sigma N \rightarrow \Lambda N$ channel coupling effect must be included 
in addition to this effective Pauli blocking,  
in a way to reproduce the relative repulsive energy shift in the 
single-particle energies for the $\Sigma$.  
This issue will be discussed in the next Section. 
The fitted value for the constant $\lambda_\Lambda$ is 
$\lambda_\Lambda = 60.25$ MeV (fm)$^3$. 
Then for the $\Sigma$ and $\Xi$ hypernuclei, we can deduce 
the constants, $\lambda_{\Sigma,\Xi}$, 
corresponding to the effective Pauli blocking effect as,  
$\lambda_\Sigma = \lambda_\Lambda$, 
and $\lambda_\Xi = \frac{1}{2} \lambda_\Lambda$,  
by counting the total number of u and d quarks in those hyperons.

%
\subsection{The effect of channel coupling}\label{channel}
We consider here the channel coupling (strong conversion) effect 
additional to the Pauli blocking at the quark level.
It is expected that the channel couplings, 
$\Sigma N - \Lambda N$ and 
$\Xi N - \Lambda \Lambda$, generally exist in hypernuclei,   
and the former coupling is considered to be especially 
important~\cite{dov,joh,dovy,mor,hau,ose,afn,afn2,aka,yam}.

First, we consider the $\Sigma N - \Lambda N$ channel coupling. 
We estimate this effect using the Nijmegen potential~\cite{nij} 
as follows. Including the effective Pauli blocking potential, 
$\lambda_\Sigma \rho_B(r)$ 
($\lambda_\Sigma = \lambda_\Lambda = 60.25$ MeV (fm)$^3$), 
we obtain as the single-particle energy for the 
$1s_{1/2}$ level in $^{\Sigma^0}_{209}$Pb, -26.9 MeV. 
This value does not contain the effect of the correct channel coupling to 
the $\Lambda$. On the other hand, in a conventional first-order
Brueckner calculation based on the standard choice of the single-particle
potentials (cf. ref. \cite{nmc} for details) the  
binding energy for the $\Sigma$ in nuclear matter with 
the Nijmegen potential~\cite{nij} is, 12.6 MeV, for the case 
without including the correct channel 
coupling effect to the $\Lambda$, namely, without the Pauli-projector 
in the $\Lambda N$ channel. When the channel coupling to the $\Lambda N$
is recovered and the Pauli-projector in the $\Lambda N$ channel is included,  
the binding energy for the $\Sigma$ in nuclear matter decreases to 5.3 MeV.
Then, the decrease in the calculated binding energy for the $\Sigma$,
12.6 - 5.3 = + 7.3 MeV, should be 
considered as a net effect of the $\Sigma N - \Lambda N$ channel 
coupling for the $\Sigma$.
In the present study, we include the effect by assuming the same form as that 
was applied for the effective Pauli blocking via $\lambda_\Sigma \rho_B(r)$, 
and adjust the parameter $\lambda_\Sigma = \lambda_\Lambda$ to 
$\tilde{\lambda}_\Sigma \neq \lambda_\Sigma$ to reproduce 
this difference in the 
single-particle energy for the $1s_{1/2}$ in $^{\Sigma^0}_{209}$Pb, 
namely, $-19.6 = -26.9 + 7.3$ MeV. Here we should point out that 
our previous study~\cite{finite1} shows that the baryon density 
around the center of the $^{208}$Pb nucleus is 
consistently close to that of nuclear matter within the model.
The value obtained for $\tilde{\lambda}_\Sigma$ in this way is, 
$\tilde{\lambda}_\Sigma$ = 110.6 MeV (fm)$^3$.

As for the $\Xi N - \Lambda \Lambda$ channel coupling, a study of 
Carr, Afnan and Gibson~\cite{afn2} shows that the effect  
is very small for the calculated 
binding energy for $^6_{\Lambda \Lambda}$He. 
The difference in the binding energy is typically less than 1 MeV  
for the results calculated with and without inclusion of 
the channel coupling effect.
Although their estimate is not for large atomic number hypernuclei,  
nor nuclear matter, we expect that the 
$\Xi N - \Lambda \Lambda$ channel coupling effect is not 
so large for the calculation of the single-particle energies, and  
thus, we neglect it in the present study. 

\subsection{Results}
Including all the effects, the Pauli blocking and 
the $\Sigma N - \Lambda N$ channel coupling 
discussed in Sections~\ref{pauli} and \ref{channel}, 
we will present the final results in this Section.

In Tables~\ref{spep1} and~\ref{spep2}, we list the calculated 
single-particle energies for $^{17}_Y$O, $^{41}_Y$Ca, 
$^{49}_Y$Ca, $^{91}_Y$Zr and $^{209}_Y$Pb hypernuclei, 
together with the experimental data for the $\Lambda$ hypernuclei.
Unfortunately, data for the larger atomic number hypernuclei 
are limited to $\Lambda$ hypernuclei. Concerning the 
single-particle energy levels for the $\Lambda$ hypernuclei, 
the QMC model supplemented by the effective Pauli blocking effect 
employed at the hadronic level, reproduces the data reasonably well. 
The small spin-orbit splittings are still achieved, as before, 
with the QMC model alone.

It is interesting to compare the single-particle 
energies for the charged hyperons, $\Sigma^\pm$ and $\Xi^-$, and 
those of the neutral hyperons, $\Sigma^0$ and $\Xi^0$. 
The present results imply that the Coulomb force is 
important for forming (or unforming) a bound state of the 
hyperon in hypernuclei. 
This was also discussed by Yamazaki et al.~\cite{yam}, in the context of 
the light $\Sigma^-$ hypernuclei.

%
\begin{table}[htbp]
\begin{center}
\caption{Single-particle energies (in MeV)
for $^{17}_Y$O, $^{41}_Y$Ca and $^{49}_Y$Ca hypernuclei, calculated 
with the effective Pauli blocking and the $\Sigma N - \Lambda N$ channel 
coupling.
Experimental data are taken from Ref.~\protect\cite{chr}.
Spin-orbit splittings are not well determined by the experiments.} 
\label{spep1}
\begin{tabular}[t]{c|ccccccc}
\hline \hline
&$^{16}_\Lambda$O (Expt.)
&$^{17}_\Lambda$O    &$^{17}_{\Sigma^-}$O
&$^{17}_{\Sigma^0}$O &$^{17}_{\Sigma^+}$O
&$^{17}_{\Xi^-}$O    &$^{17}_{\Xi^0}$O\\
\hline \hline
$1s_{1/2}$&-12.5      &-14.1 &-17.2 &-9.6  &-3.3  &-9.9  &-4.5 \\
$1p_{3/2}$&           &-5.1  &-8.7  &-3.2  &---   &-3.4  &---  \\
$1p_{1/2}$&-2.5 ($1p$)&-5.0  &-8.0  &-2.6  &---   &-3.4  &---  \\ \\
\hline \hline
&$^{40}_\Lambda$Ca (Expt.)
&$^{41}_\Lambda$Ca    &$^{41}_{\Sigma^-}$Ca
&$^{41}_{\Sigma^0}$Ca &$^{41}_{\Sigma^+}$Ca
&$^{41}_{\Xi^-}$Ca    &$^{41}_{\Xi^0}$Ca\\
\hline \hline
$1s_{1/2}$&-20.0       &-19.5 &-23.5 &-13.4 &-4.1  &-17.0 &-8.1 \\
$1p_{3/2}$&            &-12.3 &-17.1 &-8.3  &---   &-11.2 &-3.3 \\
$1p_{1/2}$&-12.0 ($1p$)&-12.3 &-16.5 &-7.7  &---   &-11.3 &-3.4 \\
$1d_{5/2}$&            &-4.7  &-10.6 &-2.6  &---   &-5.5  &---  \\
$2s_{1/2}$&            &-3.5  &-9.3  &-1.2  &---   &-5.4  &---  \\
$1d_{3/2}$&            &-4.6  &-9.7  &-1.9  &---   &-5.6  &---  \\ \\
\hline \hline
&--- 
&$^{49}_\Lambda$Ca    &$^{49}_{\Sigma^-}$Ca
&$^{49}_{\Sigma^0}$Ca &$^{49}_{\Sigma^+}$Ca
&$^{49}_{\Xi^-}$Ca    &$^{49}_{\Xi^0}$Ca\\ 
\hline \hline
$1s_{1/2}$&           &-21.0 &-19.3 &-14.6 &-11.5 &-14.7 &-12.0\\
$1p_{3/2}$&           &-13.9 &-11.4 &-9.4  &-7.5  &-8.7  &-7.4 \\
$1p_{1/2}$&           &-13.8 &-10.9 &-8.9  &-7.0  &-8.8  &-7.4 \\
$1d_{5/2}$&           &-6.5  &-5.8  &-3.8  &-2.0  &-3.8  &-2.1 \\
$2s_{1/2}$&           &-5.4  &-6.7  &-2.6  &---   &-4.6  &-1.1 \\
$1d_{3/2}$&           &-6.4  &-5.2  &-3.1  &-1.2  &-3.8  &-2.2 \\
$1f_{7/2}$&           &---   &-1.2  &---   &---   &---   &---  \\
\end{tabular}
\end{center}
\end{table}
%
%

%
\begin{table}[htbp]
\begin{center}
\caption{Single-particle energies (in MeV)
for $^{91}_Y$Zr and $^{208}_Y$Pb hypernuclei, calculated 
with the effective Pauli blocking and the $\Sigma N - \Lambda N$ channel
coupling.
Experimental data are taken from Ref.~\protect\cite{aji}.
Spin-orbit splittings are not well determined by the experiments. 
Double asterisks, $^{**}$, indicates
the value used for fitting.}
\label{spep2}
\begin{tabular}[t]{c|ccccccc}
\hline \hline
&$^{89}_\Lambda$Yb (Expt.)
&$^{91}_\Lambda$Zr    &$^{91}_{\Sigma^-}$Zr
&$^{91}_{\Sigma^0}$Zr &$^{91}_{\Sigma^+}$Zr
&$^{91}_{\Xi^-}$Zr    &$^{91}_{\Xi^0}$Zr\\ 
\hline \hline
$1s_{1/2}$&-22.5       &-23.9 &-27.3 &-16.8 &-8.1  &-22.7 &-13.3\\
$1p_{3/2}$&            &-18.4 &-20.8 &-12.7 &-5.0  &-17.4 &-9.7 \\
$1p_{1/2}$&-16.0 ($1p$)&-18.4 &-20.5 &-12.4 &-4.7  &-17.4 &-9.7 \\
$1d_{5/2}$&            &-12.3 &-15.4 &-8.1  &-0.9  &-12.3 &-5.4 \\
$2s_{1/2}$&            &-10.8 &-15.6 &-6.5  &---   &-12.4 &-3.9 \\
$1d_{3/2}$&-9.0  ($1d$)&-12.3 &-14.8 &-7.5  &-0.3  &-12.4 &-5.5 \\
$1f_{7/2}$&            &-5.9  &-10.2 &-3.1  &---   &-7.5  &-0.7 \\
$2p_{3/2}$&            &-4.2  &-10.1 &---   &---   &-7.9  &---  \\
$1f_{5/2}$&-2.0  ($1f$)&-5.8  &-9.4  &-2.3  &---   &-7.6  &-0.8 \\
$2p_{1/2}$&            &-4.1  &-9.9  &---   &---   &-7.9  &---  \\
$1g_{9/2}$&            &---   &-5.2  &---   &---   &-3.3  &---  \\ \\
\hline \hline
&$^{208}_\Lambda$Pb (Expt.)
&$^{209}_\Lambda$Pb    &$^{209}_{\Sigma^-}$Pb
&$^{209}_{\Sigma^0}$Pb &$^{209}_{\Sigma^+}$Pb
&$^{209}_{\Xi^-}$Pb    &$^{209}_{\Xi^0}$Pb\\ 
\hline \hline
$1s_{1/2}$&-27.0      &-27.0$^{**}$ &-29.7 &-19.6$^{**}$ &-10.0 &-29.0 &-19.2\\
$1p_{3/2}$&            &-23.4 &-25.9 &-16.7 &-7.7  &-25.3 &-16.3\\
$1p_{1/2}$&-22.0 ($1p$)&-23.4 &-25.8 &-16.5 &-7.5  &-25.4 &-16.3\\
$1d_{5/2}$&            &-19.1 &-22.1 &-13.3 &-4.6  &-21.6 &-12.9\\
$2s_{1/2}$&            &-17.6 &-21.7 &-12.0 &-2.1  &-21.2 &-12.0\\
$1d_{3/2}$&-17.0 ($1d$)&-19.1 &-21.8 &-13.0 &-4.2  &-21.6 &-12.9\\
$1f_{7/2}$&            &-14.4 &-18.2 &-9.5  &-0.9  &-17.6 &-9.2 \\
$2p_{3/2}$&            &-12.4 &-17.4 &-7.8  &---   &-17.1 &-8.0 \\
$1f_{5/2}$&-12.0 ($1f$)&-14.3 &-17.8 &-9.0  &-0.4  &-17.7 &-9.2 \\
$2p_{1/2}$&            &-12.4 &-17.2 &-7.6  &---   &-17.1 &-8.0 \\
$1g_{9/2}$&            &-9.3  &-14.3 &-5.5  &---   &-13.6 &-5.2 \\
$1g_{7/2}$&-7.0  ($1g$)&-9.2  &-13.6 &-4.8  &---   &-13.7 &-5.2 \\
$1h_{11/2}$&           &-3.9  &-4.9  &-1.2  &---   &-9.7  &-1.0 \\
$2d_{5/2}$&            &-7.0  &-7.5  &---   &---   &-13.3 &-3.8 \\
$2d_{3/2}$&            &-7.0  &-7.5  &---   &---   &-13.3 &-3.8 \\
$1h_{9/2}$&            &-3.8  &-4.8  &---   &---   &-9.8  &-1.0 \\
$3s_{1/2}$&            &-6.1  &-7.8  &---   &---   &-8.3  &-3.1 \\
$2f_{7/2}$&            &-1.7  &-5.8  &---   &---   &-6.2  &---  \\
$3p_{3/2}$&            &-1.0  &-3.4  &---   &---   &-6.6  &---  \\
$2f_{5/2}$&            &-1.7  &-5.8  &---   &---   &-6.3  &---  \\
$3p_{1/2}$&            &-1.0  &-3.3  &---   &---   &-6.6  &---  \\
$1i_{13/2}$&           &---   &---   &---   &---   &-3.7  &---  \\
\end{tabular}
\end{center}
\end{table}
%
%

In Table~\ref{prmst}, we show  
the calculated binding energy per baryon, 
r.m.s charge radius, $r_{ch}$ and r.m.s. radii of the hyperon, neutron 
and proton distributions, $r_Y$, $r_n$ and $r_p$, respectively.
The results listed in Table~\ref{prmst} are calculated with 
the $1s_{1/2}$ hyperon configuration for all cases.
Here, one can again notice the important role of the Coulomb force.
For example, the binding energy per baryon, 
$-E/A$, for the $\Sigma^-$ ($\Sigma^+$) 
hypernuclei is typically the largest (smallest) among the same atomic number 
hypernuclei, while the r.m.s. radii for the hyperon, $r_Y$, 
$r_{\Sigma^-}$ ($r_{\Sigma^+}$) is mostly the smallest (largest) among 
the hypernuclei of the same atomic number. 

%
%
\begin{table}[htbp]
\begin{center}
\caption{Binding energy per baryon, $-E/A$ (in MeV), r.m.s charge radius,
$r_{ch}$, and r.m.s radii of the hyperon, $r_Y$,
neutron, $r_n$, and proton, $r_p$ (in fm) for $^{17}_Y$O, $^{41}_Y$Ca, 
$^{49}_Y$Ca, $^{91}_Y$Zr and $^{209}_Y$Pb hypernuclei. 
The configuration of the hyperon, $Y$, is $1s_{1/2}$ for all hypernuclei.
For comparison, we list also the corresponding results 
for normal finite nuclei.
Double asterisks, $^{**}$, indicates
the value used for fitting.}
\label{prmst}
\begin{tabular}[t]{cccccc}
\\
\hline \hline
hypernuclei & $-E/A$ & $r_{ch}$ & $r_Y$ & $r_n$ & $r_p$ \\
\hline \hline
$^{17}_\Lambda$O     &6.37&2.84&2.49&2.59&2.72\\
$^{17}_{\Sigma^+}$O  &5.91&2.82&3.15&2.61&2.70\\
$^{17}_{\Sigma^0}$O  &6.10&2.83&2.79&2.60&2.71\\
$^{17}_{\Sigma^-}$O  &6.31&2.84&2.49&2.59&2.72\\
$^{17}_{\Xi^0}$O     &5.86&2.80&2.98&2.62&2.68\\
$^{17}_{\Xi^-}$O     &6.02&2.81&2.65&2.61&2.69\\
\hline
$^{16}$O             &5.84&2.79&--- &2.64&2.67\\
\hline \hline
$^{41}_\Lambda$Ca    &7.58&3.51&2.81&3.31&3.42\\
$^{41}_{\Sigma^+}$Ca &7.33&3.50&3.43&3.32&3.41\\
$^{41}_{\Sigma^0}$Ca &7.44&3.51&3.14&3.31&3.41\\
$^{41}_{\Sigma^-}$Ca &7.60&3.51&2.87&3.31&3.41\\
$^{41}_{\Xi^0}$Ca    &7.34&3.49&3.09&3.32&3.40\\
$^{41}_{\Xi^-}$Ca    &7.45&3.50&2.84&3.32&3.40\\
\hline
$^{40}$Ca            &7.36&3.48$^{**}$&--- &3.33&3.38\\
\hline \hline
$^{49}_\Lambda$Ca    &7.58&3.54&2.84&3.63&3.45\\
$^{49}_{\Sigma^+}$Ca &6.32&3.57&3.47&3.71&3.48\\
$^{49}_{\Sigma^0}$Ca &7.40&3.54&3.14&3.64&3.45\\
$^{49}_{\Sigma^-}$Ca &7.48&3.55&2.60&3.63&3.45\\
$^{49}_{\Xi^0}$Ca    &7.32&3.53&3.14&3.65&3.43\\
$^{49}_{\Xi^-}$Ca    &7.35&3.53&2.79&3.65&3.44\\
\hline
$^{48}$Ca            &7.27&3.52&--- &3.66&3.42\\
\hline \hline
$^{91}_\Lambda$Zr    &7.95&4.29&3.25&4.29&4.21\\
$^{91}_{\Sigma^+}$Zr &7.82&4.28&4.01&4.30&4.20\\
$^{91}_{\Sigma^0}$Zr &7.87&4.29&3.56&4.30&4.21\\
$^{91}_{\Sigma^-}$Zr &7.92&4.29&2.89&4.29&4.21\\
$^{91}_{\Xi^0}$Zr    &7.83&4.28&3.54&4.30&4.20\\
$^{91}_{\Xi^-}$Zr    &7.87&4.28&2.98&4.30&4.20\\
\hline
$^{90}$Zr            &7.79&4.27&--- &4.31&4.19\\
\hline \hline
$^{209}_\Lambda$Pb    &7.35&5.49&3.99&5.67&5.43\\
$^{209}_{\Sigma^+}$Pb &7.28&5.49&4.64&5.68&5.43\\
$^{209}_{\Sigma^0}$Pb &7.31&5.49&4.26&5.67&5.43\\
$^{209}_{\Sigma^-}$Pb &7.34&5.49&3.80&5.67&5.43\\
$^{209}_{\Xi^0}$Pb    &7.30&5.49&3.96&5.68&5.43\\
$^{209}_{\Xi^-}$Pb    &7.32&5.49&3.59&5.68&5.43\\
\hline
$^{208}$Pb            &7.25&5.49&--- &5.68&5.43\\
\end{tabular}
\end{center}
\end{table}
%

\begin{figure}[hbt]
\begin{center}
\epsfig{file=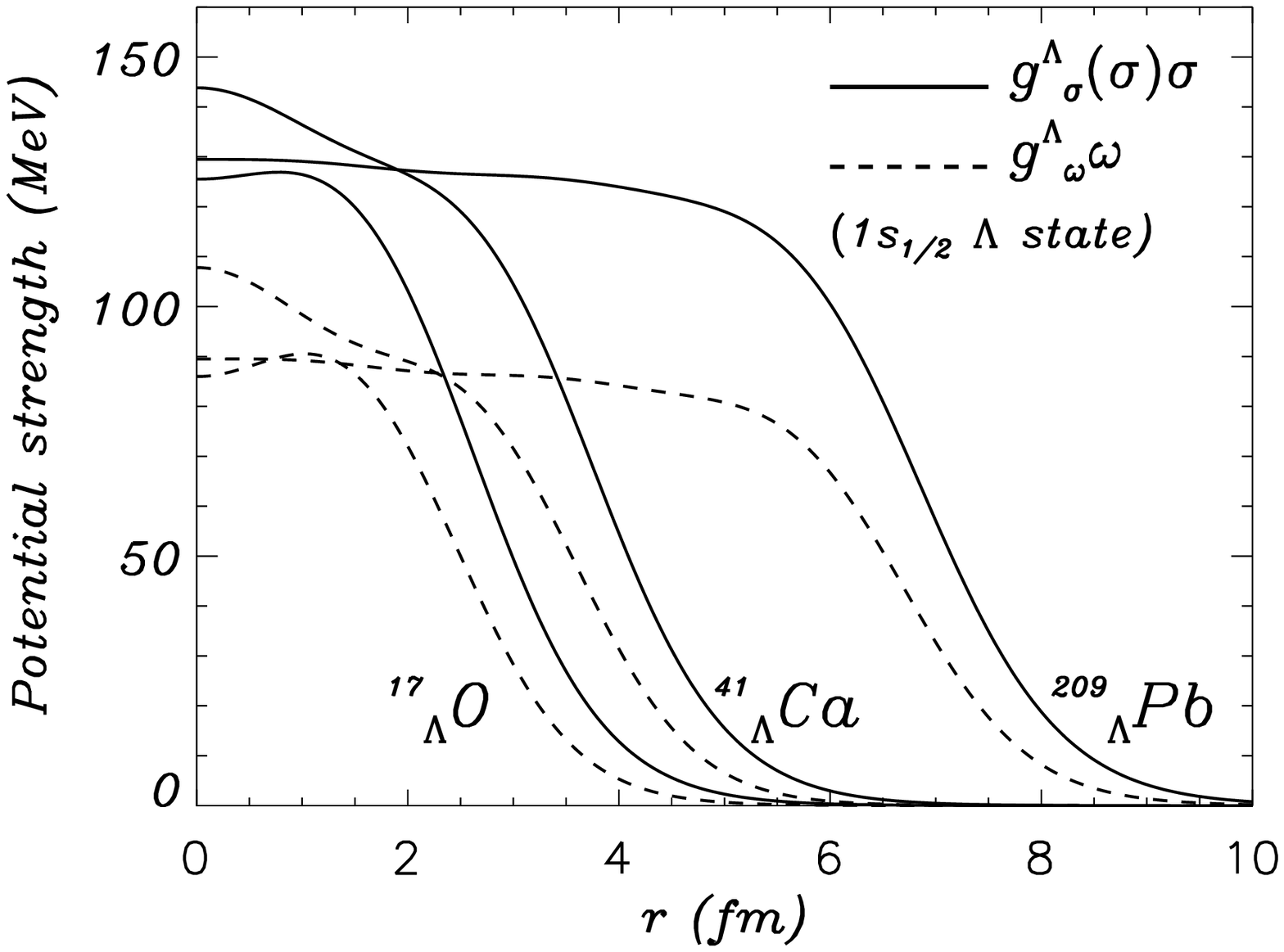,height=7cm}
\epsfig{file=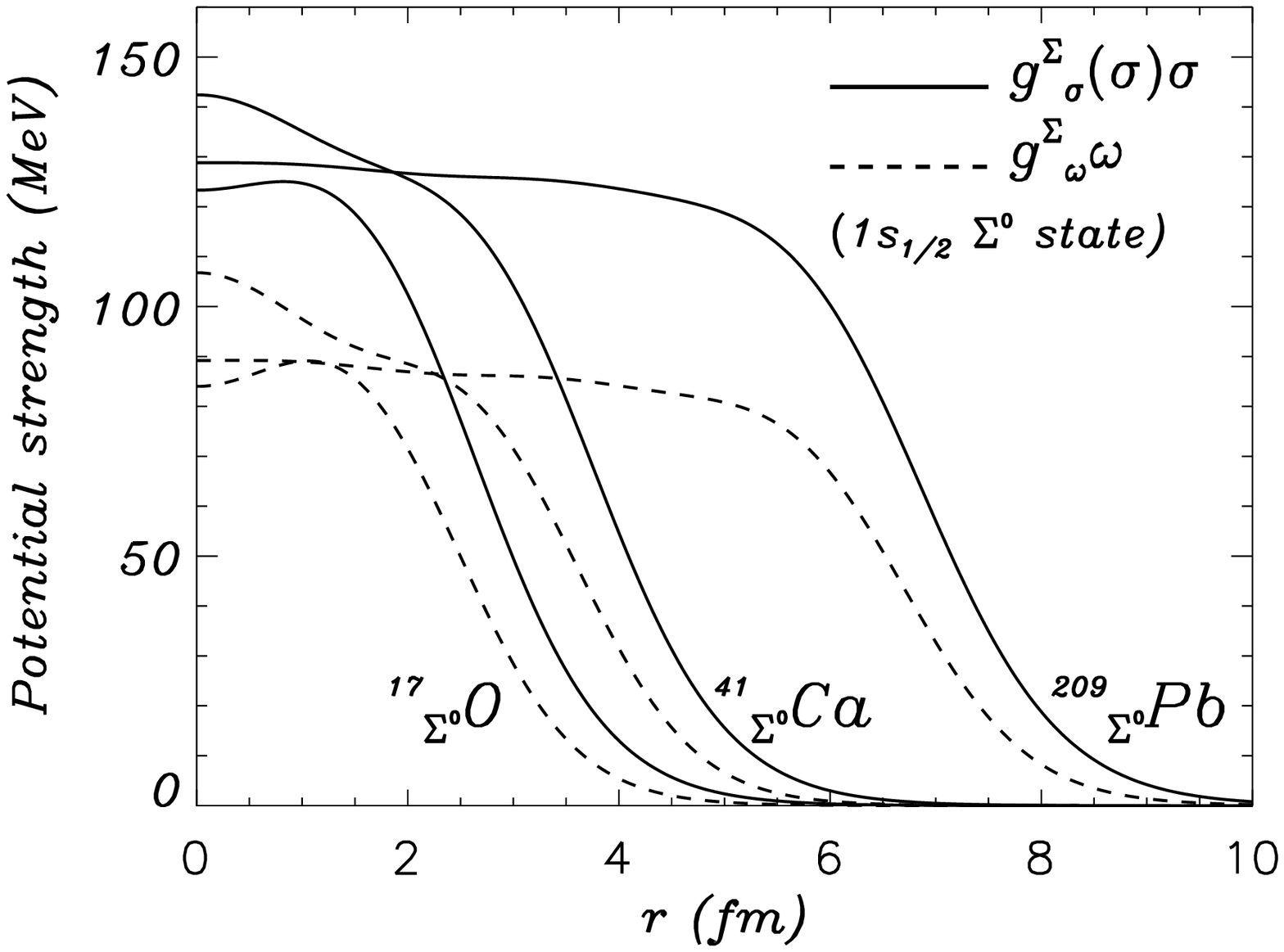,height=7cm}
\epsfig{file=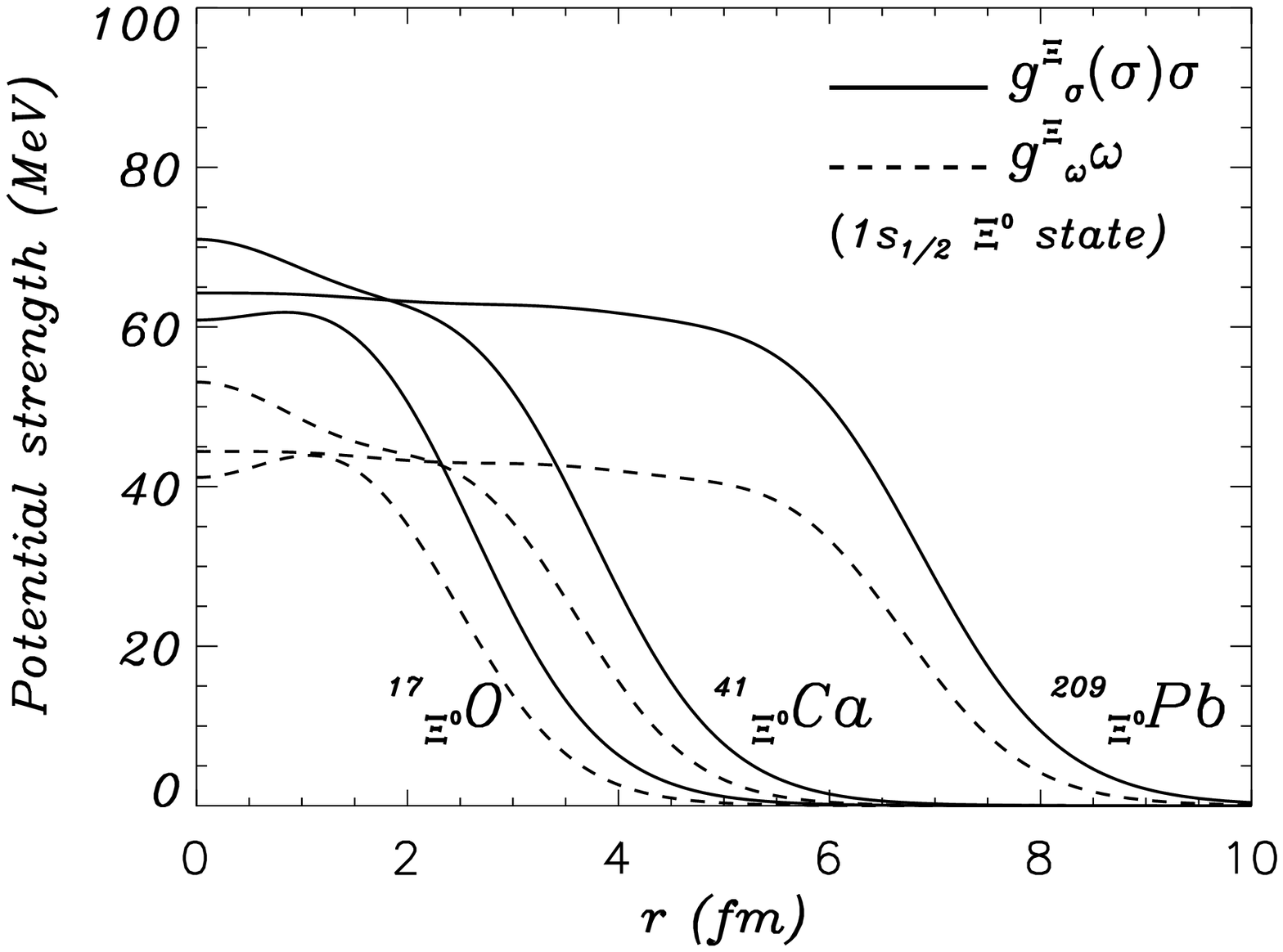,height=7cm}
\caption{Potential strengths, $g^Y_\sigma(\sigma)$ and $g^Y_\omega$, for 
$Y = \Lambda, \Sigma^0$ and $\Xi^0$ in $^{17}_Y$O, $^{41}_Y$Ca and
$^{209}_Y$Pb hypernuclei for the $1s_{1/2}$ hyperon state, calculated 
with including the effects of Pauli blocking and 
the $\Sigma N - \Lambda N$ channel coupling.}
\label{potential}
\end{center}
\end{figure}
\begin{figure}[hbt]
\begin{center}
\epsfig{file=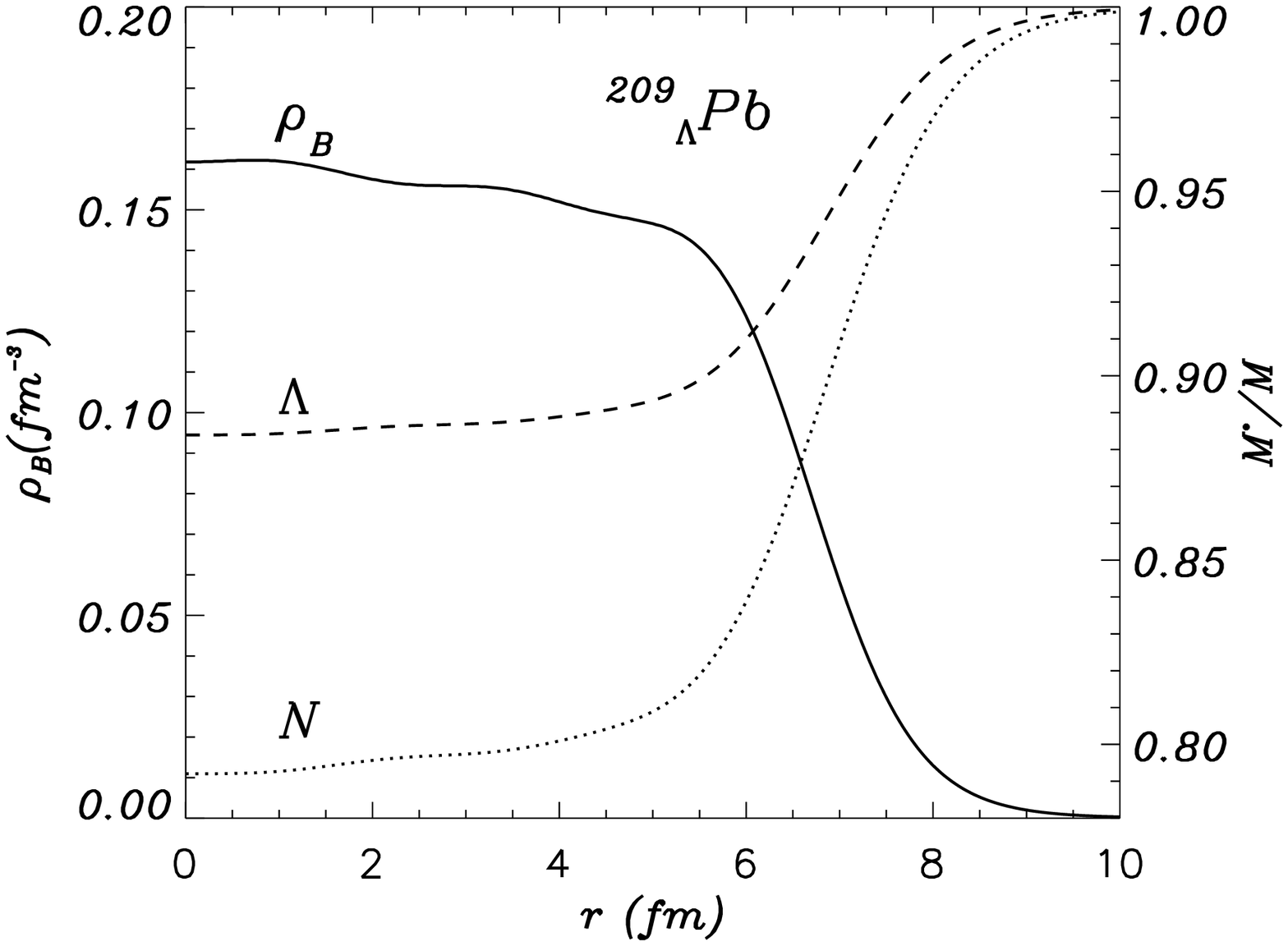,height=7.2cm}
\epsfig{file=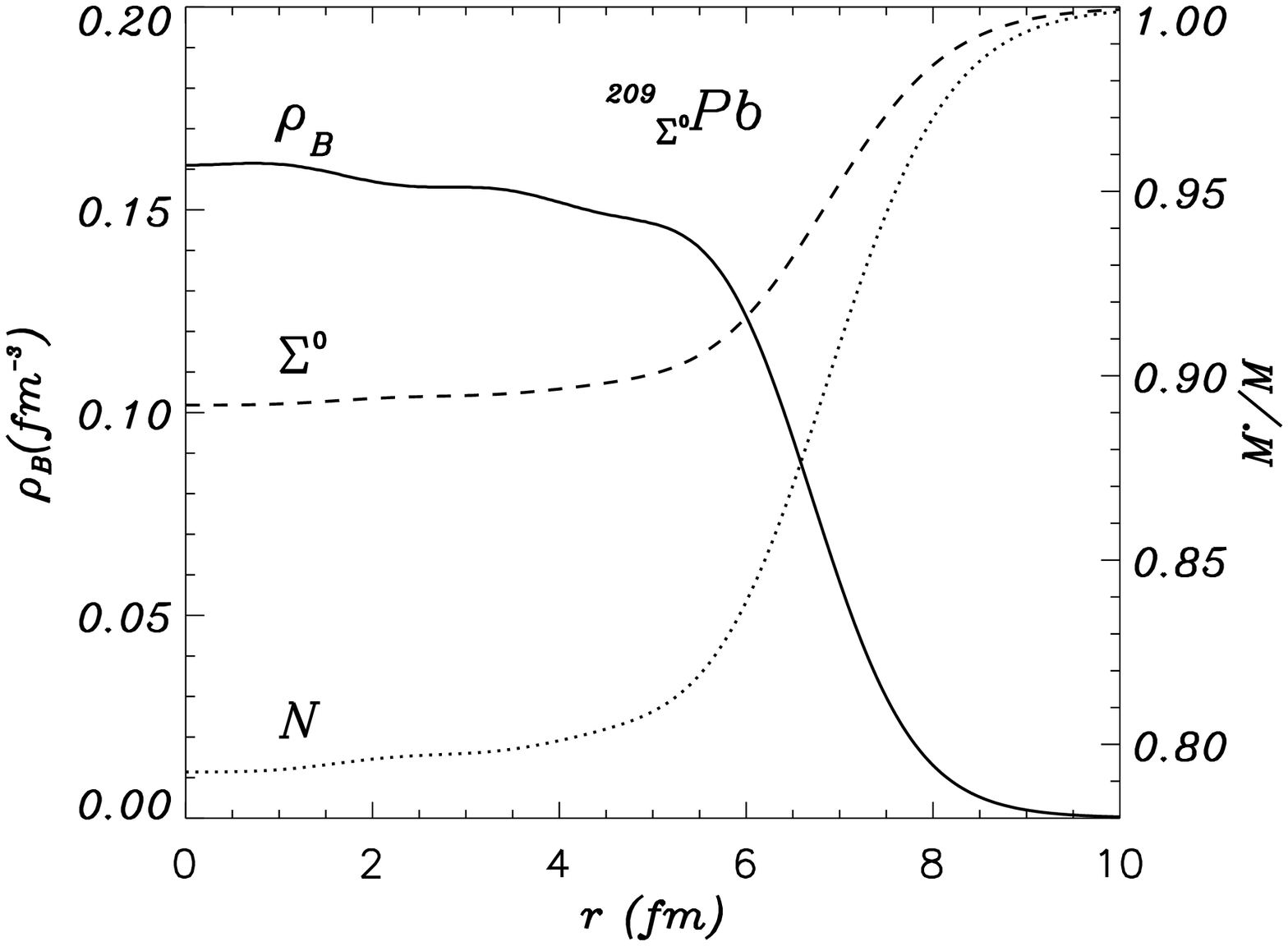,height=7.2cm}
\epsfig{file=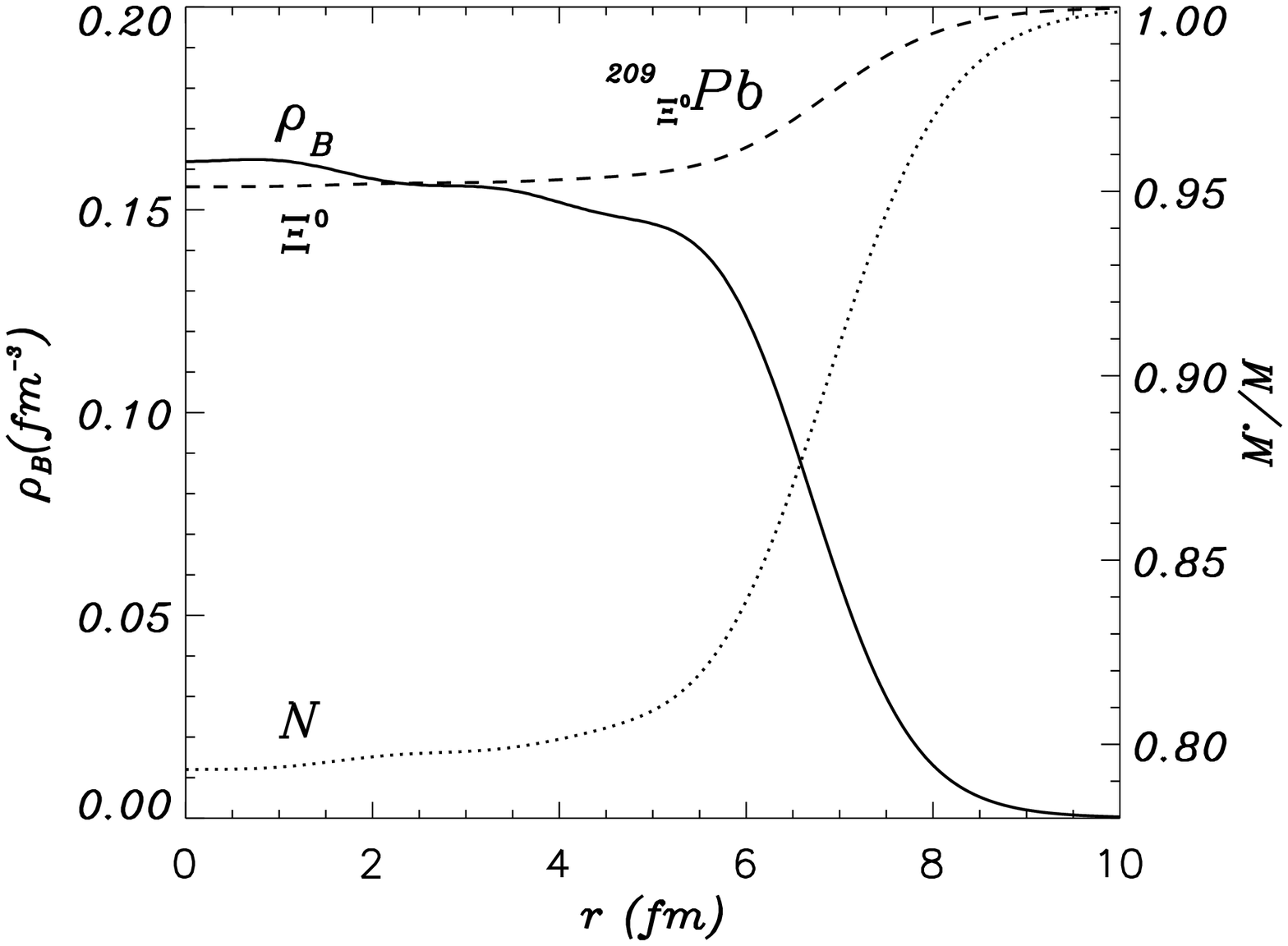,height=7.2cm}
\caption{Effective masses of the nucleon and hyperon, and the baryon 
density in $^{209}_Y$Pb hypernuclei for  
$Y$ = $\Lambda, \Sigma^0$ and $\Xi^0$.
The hyperon configuration is the $1s_{1/2}$ state for all cases.}
\label{ymass}
\end{center}
\end{figure}

Finally, we show the strengths of the scalar and vector potentials 
for the hyperons in $^{17}_Y$O, $^{41}_Y$Ca and 
$^{209}_Y$Pb  
with $Y$ = $\Lambda$, $\Sigma^0$ and $\Xi^0$ in Figs.~\ref{potential},  
and the effective masses 
of the nucleon and hyperon together with the baryon densities in 
$^{209}_\Lambda$Pb, $^{209}_{\Sigma^0}$Pb and $^{209}_{\Xi^0}$Pb 
in Figs.~\ref{ymass}.
>From the baryon densities given for each hypernucleus, one can estimate 
the magnitude of the effective 
Pauli blocking and the $\Sigma N - \Lambda N$ channel coupling effect
employed in the present study. Note that the baryon densities 
versus the distance $r$ from the center of the hypernuclei are 
slightly different for each hypernucleus due to the self-consistent 
calculations. We should also mention that the results 
shown in Figs.~\ref{ymass} are calculated by employing 
different parameters (and a different version of the model)  
from those shown in Ref.~\cite{finite2}.

%
%
%
%

%
%
%

%
%
\section{Summary and discussion}

In summary, we have reported results for 
$\Lambda, \Sigma$ and $\Xi$ hypernuclei which were systematically 
calculated with the QMC model for the first time.
Especially, the spin-orbit potentials for the hyperons in the 
hypernuclei were evaluated self-consistently with the explicit quark  
structure for the bound hyperon and nucleons.
The very small spin-orbit force for the $\Lambda$ in hypernuclei 
was achieved naturally in the present treatment.
This is a direct 
consequence of the SU(6) quark model wave function 
for the $\Lambda$ used in the QMC model.
However, the single-particle energies calculated in the QMC model alone  
tend to overestimate the experimental data.
In order to overcome this overbinding problem, 
we included the effect of Pauli blocking in a phenomenological way.
For the $\Sigma$ hypernuclei, we took into account the channel 
coupling effect estimated using the Nijmegen potential. 
In the future, these effects, which were included phenomenologically 
at the hadronic level in the 
present study, should be treated on the same footing, namely,  
at the quark level. For this purpose, it seems that 
the quark cluster model~\cite{tak,str} may 
offer some possibilities  
to include the Pauli blocking effect at the quark level. 
However, an application of the model to 
larger atomic number hypernuclei, or nuclear matter, 
seems to be difficult, and has not yet been carried out. 

Although, there seem to be several points which could be improved in 
the present treatment, we would like to emphasize that, 
this study is the first, systematic investigation of 
the properties of hypernuclei based on explicit quark degrees of freedom.
One of the advantages of the present study 
is the relatively simple nature of the 
QMC model, and simultaneously, its capability to give 
a quantitative description of the properties of finite nuclei. 

\vspace{0.5cm}

\noindent{\bf Acknowledgement}\\
We would like to thank I.R. Afnan for helpful discussions concerning 
the $\Xi N - \Lambda \Lambda$ channel coupling and letting us 
read the earlier version of the manuscript, Ref.~\cite{afn2}.
This work was supported by the Australian Research Council. 
A.W.T. and K.S. acknowledge support from the Japan Society 
for the Promotion of Science.
%
%
 \newpage

%
\end{document}